\documentstyle[11pt]{book}
\begin{document}
{\large
\begin{center}
{\Large \bf
{ BARYON AND TIME ASYMMETRIES OF THE UNIVERSE }
}\\
\vskip 1.5cm
{ Andro BARNAVELI\footnote{e-mail: bart@physics.iberiapac.ge}
and Merab GOGBERASHVILI\footnote{e-mail: gogber@physics.iberiapac.ge} } \\
\vskip 0.3cm
{\it
 {Institute of Physics of the Georgian Academy of Sciences,}\\
 {Tamarashvili str. 6, Tbilisi 380077, Republic of Georgia.}
} \\
\vskip 2.0cm
{\bf Abstract}\\
\vskip 0.5cm
\quotation
{\small

This paper is devoted to the investigation of connection between two
apparent asymmetries of the nature --- time-asymmetry and Baryon Asymmetry
of the Universe (BAU).  The brief review of this subjects is given. We
consider the particle behavior in curved space-time and the possibility
of $T$- and $CPT$-violation by the universe expansion. If these
symmetries are violated we can dispens with the nonequilibrium condition
which is usualy considered as the one of necessary ingredients for
BAU-generation.  Such mechanism of GUT-scale baryogenesis can provide
the observed value of baryon asymmetry. We show this on the example of
minimal $SU(5)$ model which usually fails to explain the obvserved BAU
withought taking into account gravitational effects. Predominance of matter
over antimatter and the cosmological arrow of time (the time-direction in
which the Universe expands) seem to be connected facts and, possibly, BAU is
the one of observable facts of $CPT$-violation in nature.

}
\endquotation
\end{center}
\tableofcontents
\chapter{Introduction}

{}~~~~~For a long time much attention was paid to the existence of
different asymmetries of nature. For example to the apparent asymmetry
between matter and antimatter and to existence of preferred time direction
in the Universe. Here we want to discuss the connections between these
asymmetries.

A striking feature of our everyday experience about time refers to its
imutable flow from past to present and from present to future. Remarkably
enough the reality of time irreversibility has been a point of
recurrent polemics in physics since the debate between Botzmann and
Zermelo late last centure (for details see \cite {Dav}). The question of
"time arrow" was discussed quite widely (see e.g. \cite{Dav,Gr,Rei,Haw,Pen}).
On the matter of irreversibility, Einstein
wrote in a letter to his friend Michelle Besso: " There is no
irreversibility in the basic laws of physics. You have to accept the idea
that subjective time with its emphasis on the now has no objective
meaning". Indeed the local physical laws which are well-known and which
people understand are all symmetric with respect to time (it would be
more correct to say that they do not change after symmetry
transformations $C$, $P$ and $T$, where $C$ interchanges particle with
its antiparticle, $P$ --- the parity transformation changes right with
left, $T$ --- changes the time direction to opposite), but nevertheless
at the macroscopic level there is an apparent time-asymmetry.

One can point out some different (at first sight) time arrows (see e.g.
\cite{Pen}):

a). Thermodynamical arrow of time, showing the time direction in which the
entropy (or disorder) does increase;

b). Psychological arrow of time. This is the direction of our "feeling" of
time --- the direction at which we remember our past but not future;

c). The cosmological arrow of time --- the time direction in which the
Universe expands but not collapses.

To this three cases one can add also:

--- The decay of $K^0$-meson (see also \cite{GyPo,TDL}), where the
$T$-violating component is only the $10^{-9}$-th part of $T$-conserving
component; here $T$-violation is derived from $CP$-violation $(\sim
10^{-9})$ under the assumption that $CPT$-violation, if such exists, is
sufficiently smaller $(\ll 10^{-9})$;

--- The quantum-mechanical observations. The procedure of observation in
quantum mechanics seems to be connected with some non-reversible process
and assumes the entropy increase;

--- The delay of radiation. Any perturbation is expanding away from its
source in all directions and reaches some distant point after particular
time-interval. The reversed picture, when perturbations are gathering from
the whole space to the source-point, seemingly, never does occur;

--- Correlation of black and white holes. General Relativity is a
time-symmetric theory.  Thus to every time-asymmetric process, such as
the formation of black holes, must correspond the other process with the
opposite time-behavior, such as would be formation of white holes. There
are different opinions about existence of white holes (see e.g.
\cite{Haw76}), but the most serious considerations lead to the conclusion
that, seemingly, such objects do not exist \cite{Pen}.

One can try to find some relations between these arrows. For example the
delay of radiation can be explained in terms of entropy increase i.e. by
the existence of thermodynamical arrow of time \cite{Pen}. Besides, basing
on the statement that the Universe has no boundaries and using the weak
anthropic principle ("we leave in such place in the Universe and
at that stage of its evolution, where our life is possible") \cite{Ant} it
is possible to show \cite{Haw} that the thermodynamical, psychological and
cosmological arrows of time have the same direction.

The psychological arrow of time is determined by thermodynamical one and
their directions coincide. One can show it on the example of computer.
When one places some information in computer's memory, he has to expend some
energy which consequently transforms into heat and thus increases the
entropy of the Universe (this growth of entropy is much greater then
its decrease which will occur since after the process of
remembering the memory of computer from the state of disorder transfers to
the state of order). Thus the time-direction in which computer does remember
is the direction in which the entropy does increase. "The entropy
increases with time because we measure the time in the direction in which
entropy increases" \cite{Haw}.

If one assumes that the Universe has no boundaries, then the well-defined
thermodynamical and cosmological arrows of time must exist, however their
directions have not to coincide inevitably. But only in the case of
coincidence of their directions the conditions for existence of
human beings could appear.  Certainly, the condition of boundary absence
means that the history of the Universe is finite but it has no
boundaries, no edges and no singularities \cite{Haw}. Then the initial
moment of time-count must be the regular point of space-time and thus the
Universe began its expansion from quite isotropic and ordered state. This
state could have some fluctuations of particle density and velocities,
but they should be sufficiently small to satisfy the uncertainty
principle. As the Universe expanded this fluctuations increased and the
Universe transferred from the isotropic and ordered state into the state
which is unisotropic and disordered. This can explain the existence of
thermodynamical  arrow of time. If the Universe finishes its expansion
and begins to collapse then disorder will continue to grow and thus the
thermodynamical and psychological arrows of time will not change their
direction in collapsing Universe. In this case thermodynamical and
cosmological arrows of time will have the opposite directions. However
here works the weak anthropic principle --- conditions at the stage of
the Universe collapse will not be valid for the existence of human
beings \cite{Haw}.  Thus the human life is possible only at the stage of
expansion and in this case the directions of cosmological,
thermodynamical and psychological arrows of time coincide.

All this arrows of time indicate that the nature does not require exact
symmetry in time. One physical consequence of this asymmetry is the decay
of $K^0$-meson. The other physical process, which can be explained by the
time asymmetry is the preferential creation of baryons over antibaryons
in the early Universe. The consequence of this process is the
predominance of matter over antimatter known as a Baryon Asymmetry of the
Universe (BAU).

BAU is the other observational asymmetry in cosmology. We all know there are
no antimatter bodies in the solar system. Solar cosmic rays are evidence
that the sun is also composed of matter. Beyond the solar system cosmic rays
prove that the asymmetry of matter and antimatter extends (at least)
throughout our galaxy.  Certainly the cosmic rays contain a $10^4$
times less amount of antiprotons than protons that is consistent with the
assumption that the antiprotons are secondaries; the flux of anti-nuclei
is less than $10^{-5}$ that of nuclei; and there is no clear detection of
an antinucleus. The clusters of galaxies either consist only of matter,
because otherwise we should observe $x$- and $\gamma $-ray emission from
matter-antimatter annihilations between the galaxies and intergalaxy gas.
There is too small information on scales larger than clusters of
galaxies. For a review see \cite{St,KT,B}.

We shall show that expansion of the Universe --- cosmological arrow of time
--- breaks Poincare-invariance and causes violation of $T$- and
$CPT$-symmetries. This fact can be expressed in such fenomena as the
different rates of direct and inverse decays of particles or mass difference
between particles and antiparticles. As a consequence of the latter the
particles in the early Universe could be created more preferentialy then
antiparticles and this had caused generation of Baryon Asymmetry of the
Universe (BAU). Thus BAU and cosmological time-arrow are connected
fenomena.  Moreover, the sign of BAU and the direction of cosmological arrow
of time are strongly correlated. We will see that in collapsing Universe
the predominance of antimatter will occur.

In section 2 we discuss the problems of particle theory in curved
space-time, in particular, in Friedmann-Robertson-Walker (FRW) space. In
subsection 2.1 we review briefly the standard cosmological model of
expanding Universe. FRW metric is conformally equivalent to
Minkowski metric and so in subsection 2.2 the formulae of conformal
transformations are given. In subsections 2.3 and 2.4 the problems of
quantization and corpuscular interpretation of quantum fields are considered
on the example of scalar field.

The section 3 is dedicated to investigation of possible violation of
$CPT$-theorem in FRW spaces. In subsection 3.1 the $CPT$-theorem in
Minkowski space is recalled briefly. Then we consider the effects
of $CPT$-violation caused by Universe expansion, such as the difference
between the direct and inverse particle-decay rates (subsection 3.2) and
the mass difference between particles and antiparticles (subsection 3.3).

In the last section 4 the connection between time and baryon asymmetries of
the Universe is discussed. In subsection 4.1 we review the problem of BAU,
in subsection 4.2 the calculation of BAU in the
frames of $SU(5)$ GUT-model is given under the assumption that $T$ and
$CPT$ is violated and the nonequilibrium condition of baryogenesis is not
necessary. In subsection 4.3 we show that such model of baryogenesis really
can provide the observed value of BAU for the different regimes of the
Universe expansion.

\chapter {Particles in the expanding Universe}

{}~~~~~The interaction of particles with gravitational fields usually is
neglected with reference to smallness of gravitational constant $G$. In
the early Universe however gravitational fields have been so strong that
particle interactions generally must not be described by quantum field
theory in Minkowski space (the same logic has to be applied to particle
interactions in the vicinity of strongly gravitating objects such as
black holes, cosmic strings, monopoles or domain walls).

Gravitational effects in the particle physics have been considered in a
number of papers. Here we shall refer to
\cite{Che,Bro,Par,GM,BF,BE,Bars,Fo,Kuz,Lo-85,Lo,GK,BG1}.  Also see books
\cite{GMM,BD} and many other references therein.

As a whole, introducing the interaction of quantum fields with
gravity is one of the principle questions of the Field Theory, since we have
not yet the successful quantum theory of gravitation. The difficulty
appeares mainly due to nonlinear structure of Einstein equations and to the
fact that gravitational theory is invariant under the infinite-parameter
group of general coordinate transformations.

The simplest possible approach to investigate the influence of strong
gravitational fields upon particle transmutations is the theory of
quantized elementary particle fields propagating in a classical
curved background space-time.

The well-known problem of investigations of matter fields in curved
space-time is the problem of interpretation of quantum fields in terms of
particles and of the description of vacuum state. Particles can be
interpreted as the energy quanta. The measured  energy corresponds to the
eigenstate of Hamiltonian which must have the diagonalized form. Particles
and antiparticles are associated with those creation-annihilation operators,
which diagonalize the Hamiltonian.

In Minkowski space the corpuscular interpretation of free fields is based on
the invariance with respect to Poincare group. We can introduce
time-conserved separation of field operators into positive and negative
frequency parts $\psi^{\pm}(x)$ and describe vacuum state $|0>$ as
$$
\psi^-(x)|0> = 0 .
$$
This modes are connected with the natural orthogonal coordinate sistem
$(t,x,y,z)$ which is associated with Poincare group.
Vector $\partial /\partial t$ is the Killing vector of Minkowski space
orthogonal to spacelike hypersurfaces $t = const$, while modes
$\psi^{\pm}$ are eigenfunctions of this vector with eigenvalues
$\mp i\omega$:
$$
\frac{\partial\psi^{\pm}(x)}{\partial t} = \mp i\omega\psi^{\pm} .
$$
 Such construction of Fock space does not depend on the choice
of the basis in the space of classical solutions of field equations.

In general case of Riemannian spaces one has no such principle to choose the
basic functions and describe the vacuum. The Poincare group is not a
symmetry group of Riemannian space and so the Killing vectors, by means of
which one can define the positive- and negative-frequency solutions, do
not exist. The mixing of positive and negative frequencies takes
place. Already for the case of free fields which are almost trivial in
the flat space this can be exhibited by particle creation from vacuum.
Moreover, if matter fields are mutually interacting in presence of a strong
gravitational field, new effects can arise, such as particle production
in some interactions and violation of $CPT$-invariance which can be
exhibited in different rates of direct and inverse decays of particles or
in the mass difference between particles and antiparticles.

In this chapter we shall discuss briefly some aspects of field theory in
curved space-time wich are necessary to understand the connection between
the Universe expansion and BAU.

\section {The model of expanding Universe}

{}~~~~~Before investigating the influence of the Universe expansion on
particle interactions let us first describe the simplest big-bang model ---
the Friedmann-Robertson-Walker (FRW) Universe (see e.g. \cite{T}).

We shall perform our calculation in the background of curved space-time
metric. The gravitational field is identified with the space-time metric
$g_{\mu\nu}$, dynamics of which is described by Einstein-Hylbert action
$$
S_g = - \frac{1}{16\pi G}\cdot \int \sqrt{-g}(R + 2\Lambda ) dV ,
$$
where, for the sake of generality, we have included a cosmological constant
$\Lambda$. Here $G \sim 1/M_{Pl}^2$ is the Newton's constant
($M_{Pl} \sim 10^{19} GeV$ is the Planck mass), $R$ is the scalar curvature,
$g$ is the determinant of metric and we have chosen the signature
$(+,-,-,-)$.
The matter is generically described by a matter action
$S_m$ which provides a source for gravitational interaction, i.e. the
right-hand side of Einstein equations
$$
R_{\mu\nu} - \frac{1}{2}(R + 2\Lambda )g_{\mu\nu} = 8\pi G T_{\mu\nu} ,
$$
where
$$
T_{\mu\nu} = \frac{2}{\sqrt{-g}}\cdot \frac{\delta S_m}{\delta g^{\mu\nu}},
$$
is the energy-momentum tensor.

Under assumptions of homogenity and isotropy the metric of the universe can
be written in the Friedmann-Robertson-Walker form:
\begin{equation}
ds^2 = dt^2 - a^2(t)dl^2 ,
\label{2.1}
\end{equation}
where $t$ is proper time measured by a comowing observer, $a(t)$ is the
scale factor and
$$
dl^2 = \frac{dr^2}{1-kr^2} + r^2 (d\theta ^2 + \sin^2\theta d\varphi^2)
$$
is the interval of 3-space of constant curvature. Here $r$, $\theta $ and
$\varphi $ are comowing spherical coordinates. The constant $k$ is a measure
of curvature and its  values $k = -1$, $1$ or $0$ correspond to open, closed
or flat isotropic models. It is easy to see that if we introduce a new
variable
$$
\bar r = \frac{1}{\sqrt{k}}{\rm arcsin}(\sqrt{k}r)
$$
then $dl^2$ takes the form
$$
dl^2 = d\bar r^2 + f^2(\bar r )(d\theta^2 + \sin^2\theta d\varphi^2) ,
$$
where $f^2(\bar r )$ is ${\rm sh}^2\bar r$ for the open Universe ($k=-1$),
$\sin^2\bar r$ for the closed Universe ($k=1$) and $\bar r^2$ for the flat
Universe ($k=0$).

To solve the Einstein equations and study the matter fields in the
background of FRW metric (\ref{2.1}) it is convenient to introduce the
dimensionless "conformal time" variable $\tau$ by the relation
\begin{equation}
\tau = \int \frac{dt}{a(t)}.
\label{2.2}
\end{equation}
Then the interval (\ref{2.1}) takes the form conformal to the flat one:
\begin{equation}
ds^2 = a^2(\tau )(d\tau^2 - dl^2) ,
\label{2.3}
\end{equation}
i.e. for any value of $k$ the (\ref{2.3}) is conformally equivalent to
Minkowski metric.

The curvature tensor for metric (\ref{2.3}) has the form:
\begin{eqnarray}
R^0_0 = \frac{3}{a^4}(\dot a^2 - a\ddot a)  ;\nonumber \\
R^i_j = -\frac{1}{a^4}(2ka^2 + \dot a^2 + a\ddot a)\delta^i_j  ; \nonumber\\
R^0_i = 0 ; \quad i,j = 1,2,3 , \nonumber
\end{eqnarray}
where the overdot denotes derivative with respect to $\tau$.

For the solution of the Einstein equations the energy-momentum tensor of
matter and cosmological constant must be determined also. Usually it is
taken the perfect-fluid ansatz for the gravity source:

\begin{eqnarray}
T^0_0 = \rho ;       \nonumber \\
T^i_j = -\delta^i_j P , \nonumber
\end{eqnarray}
where $\rho$ is the total energy density of the matter and $P$ is the
isotropic pressure, and the vanishing cosmological constant $\Lambda$ is
assumed.  Then for metric (\ref{2.3}) the Einstein equations are:

\begin{equation}
H^2 = \frac{8\pi G}{3}\rho - \frac{k}{a^2} ,
\label{2.4}
\end{equation}
\begin{equation}
\dot \rho + 3H(\rho + P) = 0 .
\label{2.5}
\end{equation}
Here the Hubble parameter $H$ characterizes the intensity of gravitational
field and is determined by the scale factor $a$ in the following way:
$$
H = \frac{1}{a}\frac{da}{dt} = \frac{\dot a}{a^2} .
$$
On the present stage of the Universe evolution
$$
H \simeq 50 \div 100 \frac{km}{sec\cdot Mps} \sim 10^{-26} cm^{-1} .
$$
It is easy to find from (\ref{2.4}) and (\ref{2.5}) that
\begin{eqnarray}
\tau =  \pm \int \frac{da}{a\sqrt{K^2\rho a^2 - k}} ; \nonumber \\
3{\rm ln}a = - \int \frac{d\rho}{(\rho + P)} + const ,
\label{2.6}
\end{eqnarray}
where
\begin{equation}
K \equiv \sqrt{\frac{8\pi G}{3}}.
\label{2.7}
\end{equation}

The system (\ref{2.4}) and (\ref{2.5}) contains three unknown functions ---
$P$, $\rho $, $a$. In order to get the unique solution of (\ref{2.4}) and
(\ref{2.5}) (up to integration constants) we have to provide an equation of
state, usually written as a fixed relation between $\rho$ and $P$. For
instance, $\rho = -P$  corresponds to vacuum dominance era ($a\sim e^{Ht}$ ,
$H = K\sqrt{\rho}$); $P = \rho /3 $ gives the radiation-dominated
cosmology ($a\sim \sqrt{t}$), while $P=0$ gives the matter-dominated ($a
\sim t^{2/3}$) expansion of the present era.

In scenarios of inflation the main ingredient is a scalar field
$\phi$, the so-called inflaton, whose potential $V(\phi )$ provides
energy and pressure during an early era of vacuum dominance. Since
the Lorentz-invariance of vacuum gives that $T_{\mu\nu} \sim
g_{\mu\nu}$, the inflation provides a negative pressure equal in
magnitude to the energy density giving an accelerated expansion. While,
since the usual matter has a non-negative pressure, the radiation- and
matter-dominated models are always decelerating ($\ddot a < 0$).

Here we would like to consider the matter fields at the energies of Grand
Unification $M_{GUT}\sim 10^{15} \div 10^{16} GeV$. This corresponds to the
age of the Universe of the order of

$$  t \sim \frac{1}{M_{GUT}} ,  $$
when inflation or radiation-dominated Universe is considered usually. At
this stage of evolution the second term at the right side of the equation
(\ref{2.4}) can be neglected and one can use the flat FRW model. Now on
the base of (\ref{2.2}) and (\ref{2.6}) it is easy to write the relation
between time and scale factor:

$$
t = \int \limits_{0}^{\tau} a(\tau )d\tau =
\frac{1}{K}\int\limits_{a_0}^{a}\frac{da}{a\sqrt{\rho}} .
$$
Here the initial moments $t_0$ and $\tau_0$, corresponding to GUT energy,
are taken to be zero for the convenience and $a_0$ is the initial value of
the scale factor.

For the vacuum-dominated  model ($\rho = const$) equations (\ref{2.6})
show that the Universe is described by de Sitter expansion low:
\begin{equation}
a(t) = a_0e^{Ht} \quad or \quad a(\tau ) = \frac{a_0}{1 - a_0H\tau}
\label{2.8}
\end{equation}
with the Hubble parameter $H = K\sqrt{\rho}$.

If one considers the radiation-dominated stage ($\rho \sim a^{-4}$), then
according to (\ref{2.6}) the scale factor depends on time as

\begin{equation}
a(t) = \sqrt{a_0^2 + 2K t} \quad or \quad a(\tau ) = a_0 + K \tau .
\label{2.9}
\end{equation}

\section {Fields in conformally equivalent spaces}

{}~~~~~As we have seen in previous section, the FRW space is conformally
equivalent to Minkowski space:
$$  \bar g_{\mu\nu} = a^2(\tau ) \eta_{\mu\nu} . $$
 This feature simplifies greatly the calculations in FRW spaces. Thus it
would be worth reminding the formalism of conformal transformations.

There exist two types of transformations named as conformal. The
conformal transformations of coordinates $x_\mu \rightarrow x'_\mu $
change only the meanings of the points on some card and do not change the
geometry.

In the field theory the more useful are the conformal transformations of
metric:
\begin{equation}
g_{\mu\nu} \rightarrow \bar g_{\mu\nu} = \Omega ^2(x)\cdot g_{\mu\nu} ,
\label{2.10}
\end{equation}
which, by contrast to coordinate transformations, contract or stretches
the manifold. In (\ref{2.10}) the $\Omega (x)$ is some continuous,
nonvanishing, finite and real function.  The conformal transformations
are used, for example, in Penrose conformal diagrams, where the whole
space-time is stretched on compact manifold. The other example is
the scale-covariant theory of gravitation which provides an interesting
alternative for conventional Einstein theory \cite{Canu}.

The conformal transformation (\ref{2.10}) changes Christoffel symbols, Ricchi
tensor and the scalar curvature in the following way
\begin{eqnarray}
\Gamma^\rho_{\mu\nu} \rightarrow \bar\Gamma^\rho_{\mu\nu} =
\Gamma^\rho_{\mu\nu} + \frac{1}{\Omega}\left( \delta^\rho_\mu D_\nu \Omega
 + \delta^\rho_\nu D_\mu \Omega - g_{\mu\nu}g^{\rho\alpha}D_\alpha \Omega
 \right) , \nonumber \\
 R^\nu_\mu \rightarrow \bar R^\nu_\mu = \frac{1}{\Omega^2}R^2_\mu -
 \frac {2}{\Omega}g^{\rho\nu}D_\mu D_\rho \left( \frac{1}{\Omega}\right) +
 \frac{1}{2\Omega^4}\delta^\nu_\mu g^{\rho\sigma}D_\rho D_\sigma \left(
 \frac{1}{\Omega^2} \right) , \nonumber \\
 R \rightarrow \bar R = \frac{1}{\Omega^2}R +
  \frac{4}{\Omega^3}g^{\mu\nu}D_\mu D_\nu \Omega ~,~~~~~~~~~~~~~~~
   \label{2.11}
  \end{eqnarray}
  where $D_\mu$ is the covariant derivative with respect to initial metric
  $g_{\mu\nu}$.

  From (\ref{2.10}) and (\ref{2.11}) it is easy to derive the following
  usefull transformation for the scalar field:
  \begin{equation}
  \left( \nabla^2 + \frac{1}{6}R \right)\varphi \rightarrow
  \left( \bar \nabla^2 + \frac{1}{6}\bar R \right)\bar \varphi  =
  \frac{1}{\Omega^3}\left( \nabla^2 + \frac{1}{6}R \right)\varphi ,
  \label{2.12}
  \end{equation}
  where
$$
  \nabla^2 = \frac{1}{\sqrt{-g}} \cdot \partial \left[ \sqrt{-g} g^{\mu\nu}
  \partial _\nu \varphi \right]
$$
  and
  \begin{equation}
  \varphi \rightarrow \bar \varphi =  \frac{1}{\Omega}\varphi .
  \label{2.13}
  \end{equation}

  The conformal invariance plays the essential role in determination of
  matter field coupling with gravitation. Indeed, massless particles are
  not characterized by definite Compton wavelength $ \lambda_c = 1/m $
  and must have the same behavior in conformally equivalent Riemannian
  spaces. It means that their motion equations must be
  invariant under conformal transformations (\ref{2.10}).

  Under this assumption one has to write the equation for massless
  scalar field $\varphi$ in 4-dimensional Riemanian space as
  \begin{equation}
  \left( \nabla^2 + \frac{1}{6} R \right) \varphi = 0 .
  \label{2.14}
  \end{equation}
  Then this equation is conformally invariant. Certainly, from (\ref{2.12})
  and (\ref{2.13}) it is easy to see that from (\ref{2.14}) immediately
  followes
$$\left( \bar \nabla^2 + \frac{1}{6} \bar R \right) \bar \varphi = 0 .$$

Equation for the massless vector field $A_\mu$
$$
D^\nu F_{\mu\nu} = 0 ,
$$
where
$$
F_{\mu\nu} = D_\nu A_\mu - D_\mu A_\nu =
\partial_\nu A_\mu - \partial_\mu A_\nu ,
$$
in four dimensional space is conformally invariant without any requirement
  for variation of $A_\mu$ under conformal transformations since
  $F_{\mu\nu}$ does not change its form in curved spaces, in particular
  under conformal transformations (\ref{2.10}).

  Dirac equation for the massless spinor field $\psi$ has the form:
  $$
  i\gamma^\mu D_\mu \psi = 0,
  $$
  where
  $$
  \gamma^\mu = h^\mu_a \gamma^a
  $$
  and $h^\mu_a$ is a tetrad gravitational field. This equation is
  conformally invariant if we require that under  transformation (\ref{2.10})
  the field $\psi$ varies in the following way:
  $$
  \psi \rightarrow \bar \psi = \frac{1}{\Omega^{3/2}}\psi .
  $$

  The fact that FRW metric and Minkowski metric are conformally
  equivalent gives us the opportunity to write the fields in cosmological
  spaces by using their form in Minkowski space. For example, for the
  scalar fields the account of the space-time curvature is equal to
  the transformation  of field

  $$ \varphi \rightarrow a(\tau )\cdot \varphi $$
  and of the Hamiltonian \cite{GMM}

  \begin{equation}
  {\cal H}^0_\varphi \rightarrow {\cal H}_\varphi =
  a^2(\tau )\cdot {\cal H}_\varphi^0 ,
  \label{2.15}
  \end{equation}
  where ${\cal H}^0$ is Hamiltonian in Minkowski space.

  The vector fields are conformally invariant and their Hamiltonian does not
  change its form in FRW space.

  For the spinor fields in FRW space-time we have

  $$ \psi \rightarrow a^{3/2}(\tau ) \psi $$
  and for their Hamiltonian \cite{GMM}
\begin{equation}
  {\cal H}_\psi = a^3(\tau ) {\cal H}_\psi^0 .
  \label{2.16}
  \end{equation}

\section{Quantum fields in FRW space}

{}~~~~~For a wide class of models gravitational field can be considered as an
external classical field, because the quantum nature of gravity is essential
only for strong gravitational fields when the space-time curvature is
characterized by Planck length
$$
l_{Pl} = \frac{1}{M_{Pl}} = \sqrt{G} \sim 1.6\cdot 10^{-33}{\rm cm} ,
$$
while for the matter fields quantum effects appear at the Compton-length
distances
$$
l_C = \frac{1}{m} ,
$$
where $m$ is the particle mass.

Thus on the GUT scale, which we are considering in this paper,
($m \sim 10^{15 \div 16}GeV \ll M_{Pl} \sim 10^{19}GeV $) gravity can be
considered as a classical field and one can investigate matter fields in
the background of curved space-time.  As an example let us consider
quantization of a scalar (pseudoscalar) field $\varphi$.  For the other-type
fields the problems arising in the FRW spaces are the
similar.

For the comparison let us recall that the scalar field
$\varphi (t,\vec x) \equiv \varphi (x_\mu )$
in Minkowski space-time is described by the action
$$ S = \int \frac{1}{2} \eta^{\mu\nu} \left( \partial_\mu\varphi
\partial_\nu\varphi - m^2\varphi^2 \right)d^4x , $$
where $\eta^{\mu\nu} = (1,-1,-1,-1)$ is the metric of Minkowski space. It
satisfies the motion equation
  $$
\left( \eta^{\mu\nu} \partial_\mu\partial_\nu - m^2 \right) \varphi =
0 .
  $$
The solutions of this equation
\begin{equation}
u_{\vec q} = \frac{1}{\sqrt{2\omega_q (2\pi )^3}}\cdot e^{i\vec q \vec x -
i \omega_q t},
\label{2.17}
\end{equation}
where
  $$
\omega_q \equiv \sqrt{\vec q^2 + m^2} ,
  $$
form the orthonormalizable basis. The modes (\ref{2.17}), which are the
eigenfunctions of operator $\partial /\partial t$ are the positive-frequency
solutions with respect to $t$:
  $$
\frac{\partial}{\partial t} u_{\vec q}(t,\vec x) = -i\omega u_{\vec
q}(t,\vec x).
  $$

To quantize $\varphi$ one has to represent it as an operator satisfying
the commutation relations
  \begin{eqnarray}
\left[ \varphi (t,\vec x), \varphi (t,\vec x')\right] =
\left[ \frac{\partial}{\partial t}\varphi (t,\vec x),
\frac{\partial}{\partial t}\varphi (t,\vec x')\right] = 0; \nonumber \\
\left[ \varphi (t,\vec x), \frac{\partial }{\partial t} \varphi (t,\vec
x')\right] = i\delta (\vec x - \vec x'). \nonumber
\end{eqnarray}

Since the fields (\ref{2.17}) form the orthonormalizable basis, the field
$\varphi$ can be represented as
\begin{equation}
\varphi (t,\vec x) = \int \frac{d^3q}{\sqrt{(2\pi )^32\omega_q}} \left[
u_{\vec q}(t,\vec x)c^-(\vec q) + u_{\vec q}^\ast (t,\vec x)c^+(\vec q)
\right] \equiv  \varphi^{(+)} + \varphi ^{(-)},
\label{2.18}
\end{equation}
where operators $c(\vec q)$ satisfy the commutation relations:
\begin{eqnarray}
\left[ c^-(\vec q), c^-(\vec q')\right] =
\left[ c^+(\vec q), c^+(\vec q')\right] = 0, \nonumber \\
\left[ c^-(\vec q), c^+(\vec q')\right] = (2\pi )^32\omega_q\delta^3(\vec q
- \vec q').
\label{2.19}
\end{eqnarray}
The fields $\varphi^{(\pm)}$ are the positive- and negative-frequency
parts of $\varphi$. Operators $c^+(\vec q)$ and $c^-(\vec q)$ have the
meaning of creation- and annihilation-operators of scalar field quanta.
In Heizenberg representation the quantum states form the Hilbert space.
If one uses the Fock basis in this space, then operator $c^-(\vec q)$ acts
on the vacuum state $|0>$ of this basis as
$$c^-(\vec q)|0> = 0  .$$

The form of Hamilton operator in the representation of
creation-annihilation operators is:
\begin{equation}
{\cal H} = \int \frac{d^3q}{4(2\pi )^3} \left[ c^+(\vec q)c^-(\vec q) +
c^-(\vec q)c^+(\vec q) \right] .
\label{2.20}
\end{equation}

Now, after we have pointed out some features of scalar field theory in
Minkowski space, let us consider the field theory in FRW space.

The scalar fields in FRW space are described by Lagrange density
$$
{\cal L} = \frac{1}{2}g^{\mu\nu}\frac{\partial \varphi}{\partial x^\mu}
\frac{\partial \varphi}{\partial x^\nu} - \frac{1}{2}\left( m^2 +
\frac{R}{6}\right) \varphi^2 ,
$$
where
$$
R = -\frac{6\ddot a}{a^3}
$$
is the scalar curvature corresponding to metric (\ref{2.3}). Here
$\ddot a$ denotes the second derivative with respect to $\tau$. The factor
$1/6$ shows that the conformal coupling to gravity is chosen (see
the previos subsection). The motion equation derived from this Lagrangian
has the form:
$$
\frac{1}{a^4}\frac{\partial}{\partial\tau}\left(
a^2\frac{\partial\varphi}{\partial\tau}\right) -
\frac{1}{a^2}\Delta\varphi + \left( m^2 - \frac{\ddot
a}{a^3}\right)\varphi = 0.
$$
The solutions of this equation can be written as
\begin{equation}
\varphi (\tau , \vec x) = \frac{1}{(2\pi )^{3/2}a} \int u(\tau , \vec p)
e^{i\vec p\vec x} d\vec p .
\label{2.21}
\end{equation}
If one redefines the field as
$$
\tilde \varphi = a\varphi,
$$
then
\begin{equation}
\ddot {\tilde \varphi} - \Delta\tilde \varphi - m^2a^2(\tau ) \tilde
\varphi = 0,
\label{2.22}
\end{equation}
i.e. one yields the Klein-Gordon equation with the time-dependent mass.
Substituting (\ref{2.21}) into (\ref{2.22}) one obtains for
$u(\tau ,\vec p)$:
\begin{equation}
\ddot u + \omega^2(\tau ) u = 0.
\label{2.23}
\end{equation}
where
$$
\omega^2(\tau ) = p^2 + a^2(\tau )m^2 .
$$
Equation (\ref{2.23}) reminds us the classical equation of harmonic
oscillator with time-dependent frequency.

For the initial conditions
\begin{eqnarray}
u^\pm (\tau_0) = \frac{1}{\sqrt{\omega_0}}, \nonumber \\
\dot u^\pm (\tau_0) = \pm i\sqrt{\omega_0}, \nonumber \\
\vec p^2 + m^2a^2(\tau_0) = \omega_0^2
\label{2.24}
\end{eqnarray}
equation (\ref{2.23}) has a solution in the integral form \cite{Bro} :
\begin{equation}
u^\pm (\tau ,\vec p) = \frac{1}{\sqrt{\omega_0}}e^{\pm
i\omega_0\tau} + \frac{m^2}{\omega_0} \int\limits_0^\tau d\tau '
u^\pm (\tau ',\vec p)V_m(\tau ')\sin (\omega_0 (\tau - \tau ')) ,
\label{2.25}
\end{equation}
where
$$
V_m(\tau ) = m^2 \left[ a^2(0) - a^2(\tau )\right]
$$
and for the simplicity $\tau_0 = 0$ is taken.
Note, that in (\ref{2.24}) and (\ref{2.25}) $\tau_0 = 0$ means the
time-moment, when the processes under investigation begin and $u^\pm$ are
the positive (negative) frequency solutions of (\ref{2.23}).

Quantization of field $\varphi $ proceeds via establishing the
commutation relations on the spacelike hypersurfaces \cite{Bro,GM}
\begin{eqnarray}
\left[ \varphi (y),\varphi (y')\right] = 0, \nonumber \\
\left[ \partial _{\mu}\varphi (y)d\sigma^\mu (y),
\partial_{\nu}\varphi (y')d\sigma^\nu (y') \right] = 0, \nonumber \\
\int_\Sigma \left[ \varphi (y), \partial_{\mu}\varphi (y') \right]
f(y')d\sigma^\mu (y') = i f(y) ,
\label{2.26}
\end{eqnarray}
where $y$ is some point on hypersurface $\Sigma$. Then introducing
operators of creation and annihilation $c^+(\vec p), c^-(\vec p)$ with
the usual commutation relations (\ref{2.19}) one can write the solution
(\ref{2.21}) in the form \cite{GM}:

\begin{eqnarray}
\varphi (\tau , \vec x) = \frac{1}{(2\pi )^{3/2}a\sqrt{2}}
\int d\vec p \frac{e^{i\vec p\vec x}}{\sqrt{1-|\lambda (\tau , p)|^2}}
\times \nonumber \\ \times
\left[\left( u^+(\tau ,\vec p) + \lambda (\tau , p)u^-(\tau ,\vec
p)\right) c^+(\vec p) +
\left( u^-(\tau ,-\vec p) + \lambda^\ast (\tau , p)u^+(\tau ,-\vec
p)\right) c^-(\vec p) \right] .
\label{2.27}
\end{eqnarray}
Here the functions $u^+(\tau ,\vec p), u^-(\tau ,\vec p)$ satisfy
(\ref{2.25}), while $|\lambda (\tau , p)|$ is an arbitrary function with
the only restriction $|\lambda (\tau , p)| < 1.$ This function $\lambda
(\tau , p) \neq 0$ is the new feature of quantization in the spaces with
nonstationary metric. If one compares equations (\ref{2.27}) and
(\ref{2.18}), one notices that the terms with $\lambda$ mix the positive-
and negative-frequency parts.  This is connected with the fact that in
such spaces energy is not conserved. In the Minkowski space-time the
requirement of positiveness of Hamiltonian of the free field with nonzero
mass leeds to unique choice of Fock representation \cite{GM}. However in
the space-time with time-dependent metric there is no energy as a
generator of symmetry group, since there is no invariance in respect to
time translations, and thus in such spaces the choice of representation
of commutation relations (\ref{2.26}) is ambiguous. It means that the
choice of vacuum, as a state invariant under available symmetry group is
also non-unique.

If one writes the Hamiltonian of such system by means of operators of
creation and annihilation \cite{GM}, then it will contain also the terms
of type $c^+(\vec p)c^+(-\vec p)$, $ c^-(\vec p)c^-(-\vec p)$
(compare with (\ref{2.20})).  This means that
Hamiltonian is not Hermittean operator. Even if $\lambda = 0$, what
corresponds to usual Fock representation, Hamiltonian has no sense of the
operator in Hilbert space. In this case the definition of particle as a
quantum of energy has no sense, since nondiagonal Hamiltonian has no
eigenvalues. (Only if $a^2(\tau ) = const$ the "dangerous" terms vanish
and Hamiltonian becomes Hermittean operator).  The solution of this
problem is that one must take $\lambda (\tau ,p) \neq 0$ and choose it in
such a way, that Hamiltonian shall have the diagonal form and "dangerous"
terms will be cancelled. The choice $\lambda \neq 0$ corresponds to the
choice of new orthonormalizable basis of solutions $\tilde u$ and to
Bogolubov transformations of creation and annihilation operators
\cite{GM}.

$$
c^-(\vec p) = \alpha_\lambda \tilde c^-(\vec p) + \beta_\lambda
\tilde c^+(-\vec p) ; $$
$$
c^+(\vec p) = \alpha^\ast_\lambda \tilde c^+(\vec p) + \beta^\ast_\lambda
\tilde c^-(-\vec p) ,$$
where
$$ \alpha_\lambda \equiv \frac{1}{\sqrt{1-\lambda^2}}; \quad
   \beta_\lambda \equiv \frac{\lambda}{\sqrt{1-\lambda^2}} . $$

Since $\lambda (\tau , p)$ is time-dependent, for each time-moment
the particular Hamiltonian is determined as a Hermittean operator in
Hylbert space where the corresponding commutation relations are
established. To different time-moments correspond different $\lambda
(\tau ,p)$ and different vacua.

\section{Short-time-interval approximation}

{}~~~~~As we have seen, in the curved space-time, in general, there is no
natural definition of what are the particles since particle creation
caused by cosmological expansion takes place.  However since the field
theory works in Minkowski space quite successfully, there must be some
approximation in the curved space-time, where the definition of particles
has a sense.  There must be the method how to choose those exact
modes of motion equation solutions which in some sence are the mostly
close to the corresponding solutions in the limit of Minkowski
space. Physically this means the construction of such minimal
disturbation of field in the expanding Universe, when the particles are
defined so that the particle creation due caused by changing of geometry
is minimal.

For example, one can investigate some process during sufficiently short
time interval when the geometry of space-time does not change
significantly. Then we can assume that $\lambda$ is constant and
construct the corresponding Hamiltonian.

 In the short-time-interval approximation
$$ a^2(\tau ) \simeq 1 + 2\dot a(0)\tau = 1 + 2H\tau ,$$
where it is taken $a(0) = 1$ and $H$ is Hubble constant. The correct
choice of $\lambda$  from (\ref{2.27}) which diagonalizes the
Hamiltonian gives \cite{GM}
$$
\lambda (\tau ,p) = \frac{m^2H}{2\omega 3}e^{i\omega \tau}\sin \omega\tau
{}.
$$
 It is clear that $\lambda \equiv 0$ when $H=0$.

The Hamiltonian depends on Hubble parameter and in the approximation with
zero Hubble parameter one yields the usual field theory in Minkowski
space.

Field quantization in FRW space leeds to some physical consequences.

In particular:

1. Since $\lambda (\tau ,p)$ is complex, $T$-invariance is
violated in corresponding field theory.

2. If in some moment the particles were described by functions
$\varphi (\tau ,\vec x)$ so that $\varphi^-(\tau ,\vec x)|0> =0$, where
$|0>$ is vacuum state, then in the next moment the "anomalous vacuum
expectation values" in respect to new vacuum $|0'>$ can appear
\cite{GM} :
$$<0'|\varphi^-(\tau ',\vec x)\varphi^-(\tau ',\vec x)|0'> \neq 0,
\quad
<0'|\varphi^+(\tau ',\vec x)\varphi^+(\tau ',\vec x)|0'> \neq 0. $$
This violates gauge invariance and thus the charge conservation. This
effect violates the conservation of any charge (among them the Baryon
charge) \cite{GM}. The value of this violation can be estimated :
$$<0'|\varphi^+\varphi^+|0'> =
(<0'|\varphi^-\varphi^-|0'>)^\ast \leq
\frac{mH}{8(2\pi )^3}e^{im\tau }\sin m\tau .$$
However one has to note that $CP$-violation caused by the Universe
expansion occurs only if Bogolubov transformations affect particles but
not antiparticles. This means that only matter but not antimatter "feels"
the expansion of the universe and antiparticles do not interact with
external gravitational field. Such model was discussed in \cite{Gri}.

3. In the FRW Universe the nonzero classical density of particle number can
appear:
  $$n_0 = i<0'|\varphi^+\dot \varphi^- - \dot \varphi^+\varphi^-|0'>
= \frac{mH^2}{16(2\pi )^3}\sin^2m\tau \simeq 10^{-46}cm^3. $$
This fenomena inspite of its small value could play essential role at the
early stages of Universe expansion.

  Now let us show that in the short-time-interval approximation the
definition of particles can have the usual meaning.

The formal solution of equation (\ref{2.23}), which reminds us the
classical equation of harmonic oscillator with time-dependent
frequency, can be written also in the form
$$
u(\tau ) = \frac{1}{\sqrt{2W}}\cdot e^{-\int_{\tau }W(\xi )d\xi}, $$
where the function $W$ satisfies the nonlinear equation
$$
W^2(\tau ) = \omega^2(\tau ) - \frac{1}{2}\left( \frac{\ddot W}{W} -
\frac{3}{2} \frac{\dot W^2}{W^2} \right) .
$$
One can expand $W(\tau )$ into series in terms of adiabatic
parameter of the oscillator
$$ \epsilon = \frac{\dot \omega}{\omega^2} = \frac{m^2}{\omega^3}\cdot
a\dot a .  $$
The zero order adiabatic approximation will be
$$
W^0(\tau ) \simeq \omega(\tau ) ,
$$
which leeds to standard modes of Minkowski space.

The maximal value for $\epsilon$ when the momentum $p = 0$ is
$$ \epsilon_{max} = \frac{\dot a}{a^2m} = \frac{H}{m} .  $$
The effect of particle creation is significant when $\epsilon\sim 1$.

If we consider the GUT scale, when $a \sim 1/{M_X}$ ($M_X$ is the
$X$-boson mass), and take a short interval of time
\begin{equation}
t \sim \frac{1}{10M_X} , \quad
\tau \sim \frac{1}{10} ,
\end{equation}
we can decompose $\omega(\tau )$ into series in terms of $\tau $:
$$
\omega = \sqrt{p^2 + a^2(\tau )m^2} \simeq \omega_0(1 +
\epsilon \omega_0\tau ) , $$
where $\omega_0 = \sqrt{p^2 + a^2_0m^2} $ and
$$\epsilon = \frac{a_0\dot a_0m^2}{\omega_0^3} .$$
One can show that adiabatic parameter is of the order of
$$\epsilon \sim \frac{M_X}{M_{Pl}} \sim 10^{-4},$$
so we can neglect particle creation and such decomposition is valid.
Certainly, for de Sitter expansion law (see (\ref{2.8}))
\begin{eqnarray}
 a = \frac{a_0}{1-K\sqrt{\rho} a_0\tau} ; \nonumber \\
 \dot a_0 = a_0K\sqrt{\rho} ,
\label{2.28}
\end{eqnarray}
where the constant $K$ is described by formula (\ref{2.7}),
we have $\sqrt{\rho } \sim M_X^2$, $\sqrt{G} \sim 1/{M_{Pl}}$ and
$$\epsilon \sim {\sqrt{G\rho}}/M_X \sim {M_X}/{M_{Pl}} .$$

For the power-function low expansion (\ref{2.9})
\begin{eqnarray}
 a= a_0 + K\tau ; \nonumber \\
 \dot a_0 = K ,
\label{2.29}
\end{eqnarray}
we have
$$
\epsilon \sim \frac{\sqrt{G}}{a_0^2M_X} \sim \frac{M_X}{M_{Pl}}
$$
again.

In the short-time-interval approximation solution of (\ref{2.22})
has the form
$$
\varphi = \left( \frac{1}{\sqrt{2}\omega_0} -
\epsilon\frac{\sqrt{2\omega_0}}{8}\tau\right)
\cdot e^{i\omega_0\tau \left( 1 + \frac{\omega_0}{2} \epsilon\tau \right)
}.  $$

For such a small values of $\epsilon$ and $\tau$ in this formula one can
neglect the terms containing $\epsilon\tau$. Thus for the sufficiently
short time intervals we have the usual definition of positive- and
negative-frequency states and we can consider the usual definition of
particles by means of creation-annihilation operators.

\chapter {Effects of $CPT$-violation in FRW Universe}

{}~~~~~At present there is good evidence that each of the three discrete
symmetries $C,P$ and $T$ by itself is only approximately valid in the
particle theory. The same applies to any of the bilinear products of
this symmetries. However the triple product $CPT$ is considered to be
exact symmetry. Hensefor elementary particle physics usually regains
symmetry when we interchange particles with antiparticles, right with
left and past with future. Thus, theories of $CPT$ violation must
necessarily step outside the standard theory.

Developments in the quantum theoty of gravity have opened possible way to
theories of $CPT$ violation. Building from his results on the spectrum of
radiation from black holes, Hawking has proposed that the generalization
of quantum mechanics which encompasses gravity allows the evolution of
pure states into mixed states \cite{Haw75}. Then it was showen that any
such dynamics leads to conflict with $CPT$ conservation (e.g.
\cite{Wald}). These ideas raised the interesting possibility that one
could find observable $CPT$ violation due to a mechanism that lies not
only beyond a local quantum description but also beyond quantum mechanics
altogether. The notion that gravitation effects beyond quantum mechanics
can effect elementary particle physics is controversial. For example, in
theories which allow the evolution of pure states into mixed states there
is a serious conflict between energy-momentum conservation and locality
\cite{Ban}.

One of the basic assumptions in proving of $CPT$-theorem is the invariance
of theory under continuous Lorentz transformations. Thus in arbitrary
gravitational field $CPT$ needs not to be realized as a compulsory
symmetry.  For example, the Universe expansion violates both --- Lorentz
invariance and time-reversal and will cause the violation of $CPT$. In
this case there is no reason to assume that particles and antiparticles
inevitably must have the same masses and lifetimes.

In this section we consider the effects of $CPT$-violation in FRW spaces.

\section{$CPT$-theorem in Minkowski space}

{}~~~~~In order to understand the effects of $CPT$-violation in the expanding
Universe let us briefly discuss some consequences of $CPT$-conservation in
Minkowski space \cite{TDL,Lu,Pau}.

        In Minkowski space for any spin-$\frac{1}{2}$ field the discrete
symmetry transformation can be defined as
\begin{eqnarray}
 T\Psi (\vec x,t)T^{-1} = \eta_t\gamma_1\gamma_3\Psi (\vec x,-t) ;
\nonumber \\
 P\Psi (\vec x,t)P^{-1} = \eta_p\gamma_0\Psi (-\vec x,t) ;\\
 C\Psi_\alpha (\vec x,t)C^{-1} =
 \eta_c(\gamma_2)_{\alpha\beta}\Psi^{\dagger}_{\beta} ,  \nonumber
\label{3.1}
\end{eqnarray}
where
$\eta_t$, $\eta_p$ and $\eta_c$ are constant phase factors $|\eta| = 1$
(the last relation is written in components). For the $CPT$
transformation, operator of which is defined as
\begin{equation}
\Theta = CPT ,
\label{3.2}
\end{equation}
we obtain
\begin{equation}
 \Theta \Psi_{\alpha}(\vec x,t) \Theta^{-1} = \eta
(i\gamma_5)_{\alpha\beta}\Psi^{\dagger}_{\beta}(-\vec x,-t) ,
\label{3.3}
\end{equation}
where $\eta$ is the product of arbitrary phase factors $\eta_t$, $\eta_c$
and $\eta_p$.

For the boson fields $\varphi (x)$ we have
\begin{equation}
 \Theta \varphi (x) \Theta^{-1} = \eta \varphi^{\dagger}(-x).
\label{3.4}
\end{equation}

To demonstrate how the $CPT$ theorem is proved, let us consider a local
field theory with different spin fields and demand that the theory must
be invariant under continuous Lorentz group. All these fields satisfy the
usual spin-statistics relation: integer-spin fields obey Bose statistics,
while half-integer-spin fields obey Fermi statistics.

If we assume that the theory can be described by a local $n$-th rank
Lorentz-tensor $F_{k_1...k_n}$ which is a normal product of
fields, then considering all possible field-combinations and using
(\ref{3.3}), (\ref{3.4}) one can show that \cite{TDL}
$$\Theta F(x) \Theta^{-1} = (-1)^n \cdot F^{\dagger}(-x) .$$
The Lagrangian density $L$ is one of such products and it
is a tensor of $0$-th rank. Therefore setting $n=0$ we obtain
\begin{equation}
\Theta L(x) \Theta^{-1} = L^{\dagger}(-x) .
\label{3.5}
\end{equation}
In a quantum field theory the Lagrangian is a Hermitean operator
$L=L^{\dagger}$.  The action is the 4-dimensional volume integral $ S =
\int Ld^4x $.  Using (\ref{3.5}) we see that $S$ is invariant under $CPT$
transformation.  Thus the theory and the corresponding Hamiltonian is
invariant under $CPT$ :

\begin{equation}
\Theta {\cal H} \Theta^{-1} = {\cal H} .
\label{3.6}
\end{equation}
In proving the $CPT$-theorem the following assumptions are used: (a).
Invariance under the continuous Lorentz transformations; (b).
spin-statistical relations and (c) the locality of field theory.

Now let us consider a massive particle $\psi$ at rest. Let the state
$|\psi >_l$ be the one with its $z$-component angular momentum equal to $l$.
Apart from a multiplicative phase factor the state changes under the
$C$-operation in the following way:
$$
C|\psi >_l \rightarrow |\bar \psi >_l .
$$
Under $P$-operation it remains itself :
$$
P|\bar\psi >_l \rightarrow |\bar \psi >_l ,
$$
but under $T$, since $l$ changes its sign,
$$
T|\bar\psi >_l \rightarrow |\bar \psi >_{-l} .
$$
Therefore the $CPT$ transformation (\ref{3.2}) makes the following
changes:
\begin{equation}
\Theta |\psi >_l = e^{i\alpha}|\bar \psi >_{-l} ,
\label{3.7}
\end{equation}
where $\alpha$ is constant phase factor.

The mass of the particle $\psi$ is given by the expectation value
\begin{equation}
m = <\psi |{\cal H}|\psi >_l  ,
\label{3.8}
\end{equation}
where ${\cal H}$ is the total Hamiltonian, clearly real and independent
on $l$.  Since (\ref{3.8}) equals to its complex conjugate, one can write
$$
m = <\psi |{\cal H}|\psi >_l^* =
<\psi |\Theta^{-1}\Theta {\cal H} \Theta^{-1}\Theta |\psi >_l .  $$
Due to (\ref{3.6}) and (\ref{3.7}) the above expression can also be
written as
\begin{equation}
m = <\bar \psi |{\cal H}|\bar \psi >_{-l} = \bar m .
\label{3.9}
\end{equation}
Thus the $CPT$-invariance of Hamiltonian leeds to the mass equality between
particles and antiparticles.

The other application of $CPT$-theorem is the lifetime equality between
particles and antiparticles.

Let us consider the decay of particle $\psi$ and its antiparticle
$\bar\psi$ through some interaction Hamiltonian ${\cal H}_{int}$. To the
lowest order of ${\cal H}_{int}$ the decay-widths (the inverse
quantities to the lifetimes) of $\psi$ and $\bar\psi$ are given by the
perturbation formulae
$$
\Gamma_\psi = 2\pi \sum_f \delta (E_f - E_i) \left| <f|G(\infty ,0){\cal
H}_{int}|i> \right|^2   ;
$$
$$
\Gamma_{\bar\psi} = 2\pi \sum_{\bar f} \delta (E_{\bar f} - E_{\bar i})
\left| <{\bar f}|G(\infty ,0){\cal H}_{int}|{\bar i}> \right|^2 ,
$$
where $|i>$ and $|f>$ are the initial and the final states and $G(t,t_0)$
is Green's function of motion equation in the representation of
interaction. If $CPT$-invariance holds, then
$$
\Theta G(t,t_0) \Theta^{-1} = G(-t,-t_0) .
$$
Using this formula we can convert the above expression for $\Gamma_\psi$
into
$$
\Gamma_\psi = 2\pi \sum_f \delta (E_f - E_i) \left|
<f|\Theta^{-1}\Theta G(\infty ,0)\Theta^{-1}\Theta {\cal
H}_{int}\Theta^{-1}\Theta |i> \right|^2 =
 $$
$$
 = 2\pi \sum_{f} \delta (E_{f} - E_{i})
\left| <{\bar f}|G(-\infty ,0){\cal H}_{int}|{\bar i}> \right|^2 .
$$
Using definition of $S$-matrix
$$
G(-\infty ,0) = S^\dagger G(\infty ,0)
$$
and the relations between the initial and the final energies
$$
E_f = E_{\bar f},
$$
$$
E_i = E_{\bar i},
$$
we can write the equation of decay-width in the form
$$
\Gamma_{\psi} = 2\pi \sum_{\bar f} \delta (E_{\bar f} - E_{\bar i})
\left| <{\bar f}|S^\dagger G(\infty ,0){\cal H}_{int}|{\bar i}> \right|^2
=
$$
$$
 = 2\pi \sum_{\bar f} \delta (E_{\bar f} - E_{\bar i})
\left|\sum_{\bar f'} <{\bar f}|S^\dagger |\bar f'><\bar f'| G(\infty
,0){\cal H}_{int}|{\bar i}> \right|^2 .
 $$
Since $S^\dagger S = 1$ and the $S$-matrix has the only nonvanishing
matrix elements between the equal-energy states \cite{Mey} , we have
$$
 \sum_{\bar f} \delta (E_{\bar f} - E_{\bar i})
 <{\bar f}|S^\dagger |\bar f'><\bar f| S^\dagger |\bar f"> =
  \delta (E_{\bar f'} - E_{\bar i}) \delta_{\bar f' \bar f"} .
 $$
Using this formula it is easy to see that
$$
\Gamma_{\psi} = 2\pi \sum_{\bar f'} \delta (E_{\bar f'} - E_{\bar i})
\left| <{\bar f'}| G(\infty ,0){\cal H}_{int}|{\bar i}>
\right|^2 = \Gamma_{\bar\psi} .
$$
This establishes the lifetime equality between $\psi$ and $\bar \psi$ to
the lowest order of the decay Hamiltonian ${\cal H}_{int}$.

There are several tests which check the $CPT$ invariance. The most simple
ones are the equality of masses and lifetimes of a particles and
antiparticles \cite{Part}. The best experimental support of
$CPT$-invariance at the present stage of the Universe expansion is given
from such identity for $K$-mesons \cite{GyPo}. Existing experimental evidence
for $CPT$
invariance is poor, e.g. the limit for the strength of the
$CPT$-violating interaction is only $1/10$ of the $CP$-violating
interaction \cite{Kob}.

\section{The direct- and inverse-decay rate difference}

{}~~~~~Now let us discuss some effects of $CPT$-violation in strong
gravitational fields.

Parker \cite{Par} concerned with particle creation by the gravitational
field and its possible applications to cosmology while in \cite{BF} some
effects of interactions were considered. Attention was paid also to the
possibility of $CPT$-violation in the presence of a time-dependent
gravitational field which breaks Lorentz-invariance. Brout and Englert
\cite{BE} have discussed the possibility of $CPT$-violation induced by
the expansion of the Universe. However in the particular model which they
investigated such a violation does not in fact occur. Barshay \cite{Bars}
had also discussed the effects of a fenomenological $CPT$-violating
interaction in the early Universe.

In papers of Ford \cite{Fo}, Grib and Kryukov \cite{GK} and Lotze
\cite{Lo} the effect of expansion of the Universe upon the rate of decay
of a massive  particles was investigated. There was argued that the
average decay rate of particle in general is not $CPT$-invariant. Authors
calculated the difference between the direct and inverse decay rates in
the background of FRW universe. This difference and the consequent $CPT$
violation arises because the expansion of the Universe causes a mixing of
positive and negative frequency modes (see subsection 2.3); it is this
mixing which is responsible for both particle creation \cite{Par} and
$CPT$ violation.  Brout and Englert \cite{BE} do not find $CPT$ violation
in their models because they worked in the approximation where the mixing
of positive and negative frequencies is ignored.

Ford had derived that the difference between the rate of decay of massive
scalar particle $\Phi$ with momentum $\vec p$ into $n$ massless particles
$\varphi$ and the rate of inverse decay depends on a quantity
\begin{eqnarray}
\Delta_{CPT} = \frac{1}{\omega\cdot\Delta\tau} \left[
\int\limits_{-\infty}^{+\infty} d\tau a^{3-n}(\tau )\sin (2\omega\tau )
\cdot
\int\limits_{-\infty}^{+\infty} dy \cos (2\omega y) V(y) - \right.
\nonumber \\
\left.
\int\limits_{-\infty}^{+\infty} d\tau a^{3-n}(\tau )\cos (2\omega\tau )
\cdot
\int\limits_{-\infty}^{+\infty} dy \sin (2\omega y) V(y)  \right]
\label{3.10}
\end{eqnarray}
where
$$
\omega = \sqrt{(\vec p)^2 + m^2a^2(\pm\infty )},
$$
\begin{equation}
V(\tau ) = m^2 \left[ a^2(\pm \infty) - a^2(\tau ) \right]
\label{3.11}
\end{equation}
(compare with (\ref{2.25})) and it is assumed that
\begin{equation}
a(+\infty ) = a(-\infty ).
\label{3.12}
\end{equation}
The quantity (\ref{3.10}) is a measure of $CPT$-violation. Note that if
$a(\tau)$ is an even function or constant, then $\Delta_{CPT}$ vanishes.
Similarly, if one lets $a(\tau ) \rightarrow a(-\tau )$, then
(\ref{3.10}) changes its sign. Thus $CPT$-invariance is preserved in a
generalized sense: the average decay rate in a Universe is equal to the
average inverse decay in the Universe with the time reversed scale
factor. Note also, that (\ref{3.10}) is equal to zero if $n=3$. Thus
$CPT$ violation is possible in this model only if the interaction is not
conformally invariant.

Consequently Ford required three conditions to be fulfilled
simultaneously in order that the interaction of scalar particles
considered by him violated $CPT$-invariance:

1. The conformal scale factor must be an even function of the conformal
time parameter.

2. The conditions for particle creation in non-interacting theory have to
be fulfilled for at least one of the quantum fields participating in the
interaction.

3. Additionally, the interaction itself has to break conformal
invariance.

More generalized case of the decay of massive scalar particle $\Phi $
into $n$ massive scalar particles $\varphi $ was considered by Grib and
Kryukov \cite{GK}. They obtained that the difference between the rates of
decay and inverse decay are different in this case even if $n=3$ (however
again $a(\tau )$ must be not time-even), since in this case the
interaction is not conformally invariant.

Lotze \cite{Lo} has calculated the rates of $\pi^0 \rightarrow
\gamma\gamma$ and $\gamma \rightarrow e^+e^-$ decays and inverse decays
in the background of time-asymmetrically expanding Universe. He derived,
that $CPT$ is violated in this decays even though the interaction of
photons and fermions does not change the conformal-invariance properties
of the theory. Thus the third condition is not necessary in this case.

In all the cases the effect of $CPT$ violation is caused by the Universe
expansion (collapse), which leads to the mixing of positive and negative
frequencies at the exponents in the solutions of the field equation; i.e.
the effect of $CPT$ violation is caused by the coefficients of the
exponents due to frequency mixing. This coefficients (at the exponents
with mixed frequencies) are responsible also for particle creation by the
gravitational field. This means that Characteristic time, when the effect
is significant, is the Compton time (and not the Planck time) determined
by particle masses. One can derive that $\Delta_{CPT}$ vanishes when
$\Delta\tau\cdot\omega \rightarrow \infty$. This reflects the fact that
the wavelengths much shorter than the local radius of curvature of
space-time are not effected by the background gravitational field. I.e.
$CPT$-violation is the most noticeable when the particle's wavelength is
of the order of the radius of curvature.

It is clear that in the absence of gravitational field (i.e. in the case
of Minkowski space-time) the effect of $CPT$-violation is absent, while
in the presence of non-stationary gravitational field, which is not
invariant in respect to time-reflection (this is the case in FRW
cosmology) the effect of $CPT$-violation is nonzero. Thus the measure of
$CPT$-violation can be associated with the parameter describing intensity
of gravitational field. Such the parameter is Hubble parameter
$H = \dot a(\tau )/a^2(\tau )$, which at present is equal to $H \sim
6\cdot 10^{-27} cm^{-1}$.

Let us show, that the measure of some effect caused by $CPT$ violation
due to variation of $a(\tau )$ during the time interval $\Delta\tau$
beginning from some initial moment $\tau_0$ is proportional to $m^2H$
(where $H$ is the value of Hubble parameter for the time interval under
consideration). In particular, $\tau_0$ can be also a singular point.

For the fenomena considered in papers \cite{Fo,Lo,GK} the measure of
$CPT$ violation is determined by formula (\ref{3.10}) which includes the
quantity (\ref{3.11}). It is clear that
$V(\tau )/\Delta\tau \sim \left( a^2(\pm\infty ) -
a^2(\tau )\right) /\Delta\tau \sim \dot a(\tau )|_{\tau = c}$,
where $c$ is some moment in the considered time interval $\Delta\tau $.
Since for the models with the condition (\ref{3.12}) the quantity
$1/a^2(\tau )$ always can be evaluated by some finite number, the whole
expression (\ref{3.10}) is proportional to $m^2H$.

One can show that the measure of $CPT$-violation is proportional to
$m^2H$ also in other models including the models with singularities (see
\cite{GK}). This means that the effect is significant for Compton times.

\section{Particle-antiparticle mass-difference}

{}~~~~~The other consequence of $CPT$-violation caused by the Universe
expansion can be the mass-difference between particles and antiparticles
\cite{BG1}.

In subsection 2.2 we have seen that when the curvature of FRW metric is
taken into account via conformal transformations of field operators, the
scalar, spinor and vector fields behave in different manner. This
difference can be exhibited when one investigates the particle
interactions in FRW space. In particular, taking into account the
formulae of conformal transformations (see subsection 2.2) it is easy to
find that the mass acquired by initially massless vector particles (e.g.
leptoquarks) through the Higgs mechanism becomes time-dependent
\cite{GMM}:
\begin{equation}
M_X^2 = a^2(\tau )g<\varphi >.
\label{3.13}
\end{equation}
Here $g$ is the Yukawa coupling constant and $<\varphi >$ is the vacuum
expectation value of the Higgs field.

Here arises the problem, how to define, what is particle and what is
antiparticle, since equation (\ref{3.13}) corresponds to the classical
equation for the harmonic oscillator with time-dependent frequency
$$\omega^2(\tau ) = p^2 + M_X^2(\tau )$$
and in this case the time
reflection (opposite to flat-wave case) does no longer transfer the
eigenvalue with positive frequency into eigenvalue with negative
frequency, i.e. particle into antiparticle.

This problem can be avoided if the particles acquire their masses during
sufficiently short period
$$
\tau \ll 1.
$$
Then it is possible to use the short-time-interval approximation (see
subsection 2.4).  Let us decompose $a(\tau )$ into a series in $\tau$ and
leave only the  first order in the wave equation. For the time part of
wave  function  we obtain the flat-wave expression ${\rm
exp}(i\omega_0\tau)$, where
$$\omega_0 = (p^2 + m_X^2)^{1/2} ,$$
and
$$
m_X^2 = a_0^2<\varphi >^2 .
$$
In this approximation the frequency $\omega_0$   does  not  depend on time
and the usual definition of antiparticles (as a  particles with negative
frequency or "moving backwards" in time) and the form of $CPT$-transformation
operators (\ref{3.1}) remains valid. Corrections connected with the
time dependence of metric remain only in the expression of the
Hamiltonian of interactions (see subsection 2.2) and in the mass of
leptoquarks. In the first order of $\tau $
$$ M_X(\tau ) \sim m_X + \Delta m_X ,$$
where
\begin{equation}
\Delta m_X = (\dot a|_{\tau = 0})m_X\tau /a_0
\label{3.14}
\end{equation}
(the point means the derivative with respect to $\tau$ ).

For the mass of antileptoquarks (obtained by the time reflection $\tau
\Rightarrow -\tau$) we  have:
$$ M_{\bar X}(\tau ) \sim m_X - \Delta m_X .$$
In the expanding Universe $\Delta m_X$ is positive $(\dot a > 0)$
and the mass of leptoquarks is larger then the mass of antileptoquarks.
Below we shall see that this leads  to the predominance of matter over
antimatter.  In the collapsing  Universe $\Delta m_X$ is negative  since
$\dot a < 0 $ and we should observe the  opposite  picture.

Herefrom we can conclude that the sign of BAU and the direction of
cosmological arrow of time are correlated.

\chapter{Cosmological arrow of time and BAU}

{}~~~~~Different asymmetry features of Nature are, seemingly, connected
with each other. We shall see that Baryon Asymmetry of the Universe is
correlated with the other asymmetry --- cosmological time arrow.
Cosmological expansion causes violation of $T$- and $CPT$-symmetry, while
the latter provides the mass-difference between particles and
antiparticles.  In this case even in thermodynamical equilibrium there
can be the predominance of matter over antimatter. This means that the usual
scenario of BAU generation changes radically.

In this section we shall calculate the BAU generated on GUT scale in
thermal equilibrium when $CPT$ is violated due to the Universe expansion.

\section {The problem of baryon asymmetry}

{}~~~~~One of the basic observational facts in cosmology is the apparent
predominance of matter over antimatter.

The Baryon Asymmetry of the Universe (BAU) can be described by the
quantity
$$
\Delta (t) = \frac{n_b(t) - n_{\bar b}(t)}{n_b(t) + n_{\bar b}(t)} ,
$$
where $n_b(t)$ and $n_{\bar b}(t)$ are the densities of baryons and
antibaryons at the moment $t$. At present $( t = 3\cdot 10^{17} sec )$ this
quantity $\Delta (3\cdot 10^{17}sec) \approx 1$, since
$n_{\bar b} \approx 0$.  At the temperatures
$T > m_{nucl}$, i.e. $t < 10^{-6} sec$, the amount
of nucleon-antinucleon (quark-antiquark) pairs in plasma is equal to amount
of photons (the difference is a factor $\sim 1$). So

$$ \Delta (t \leq 10^{-6}sec)
\sim \frac{n_b - n_{\bar b}}{n_\gamma } \equiv \delta , $$
where
$n_\gamma $ is the photon density. As the Universe expands isentropycally,
the quantity $\delta $ changes with time very weakly (it changes by factor
$\sim 5$ due to photon gas heating through the heavy particle annihilation),
since at $T \sim 1GeV \sim 10^{13}$K the rate of possible B-violating
processes is negligible. Thus
$$ \delta = \frac{n_b - n_{\bar b}}{n_\gamma }
\big|_{t\leq 10^{-6}sec} \approx \frac{n_b}{n_\gamma} \big|_{t=3\cdot
10^{17}sec} .  $$

Essentially the photons in the Universe are in 3 K background; the number
density of photons is
$n_\gamma = 2\zeta (3)/(\pi^2T_\gamma^3) ,$
where $\zeta (3) \simeq 1.202$ and $T_\gamma \simeq 2.9K$.

The present baryon density $n_b$ is conveniently expressed in terms of the
parameters $h$ (Hubble parameter $H \equiv 100h\ km \cdot sec^{-1} \cdot
Mps^{-1}$ ) and $\Omega = \rho_b / \rho_c $ ($\rho_c = 3H^2/(8\pi G)$ is the
critical density of the Universe \cite{KT} and $\rho_b = m_b\cdot n_b$ is
the baryon mass density):
$$
n_b = \Omega h^2\cdot 1.13\cdot 10^{-5} cm^{-3} .
$$
This is also the net baryon number density $n_B$. Our knowledge of $h$ and
$\Omega$ is poor; we take as extreme limits $1/2 < h < 1$ and $0.005 <
\Omega < 2$. (The upper limit $\Omega = 2$ corresponds to a deceleration
parameter $q_0=1$, while galactic masses determined from rotation curves
indicate $\Omega > 0.005$). With these ranges we find that the
baryon-per-photon ratio is (see e.g. \cite{FOT}):
\begin{equation}
\delta = \frac{n_B}{n_\gamma} \sim 10^{-11} \div 10^{-8} .
\label{4.1}
\end{equation}

Very often, instead of (\ref{4.1}), the baryon number density to entropy
density ratio $n_B/s$ is considered. Approximately $n_B/s \approx 1.7
n_B/n_\gamma $ (see \cite{FOT}). The entropy $s$ per comoving volume is a
more useful quantity because $s$ is a constant for an adiabatically
expanding Universe, while, as we mentioned above, $n_\gamma$ is modified by
processes such as $e^+e^-$ annihilation.

The quantity $\delta$ is one of the fundamental quantities of the Universe.

Different kinds of models, truing to explain the origin of BAU and value of
$\delta$ were constructed (for the review see \cite{DZ,L}).

If the baryon number were conserved, the observed baryon excess would
probably have to be postulated as an initial condition on the big bang. The
requirement of such a small number as an arbitrary initial condition is
possible but very unesthetic. The other possibility, that the total baryon
number of the Universe is zero but baryons and antibaryons exist in widely
separated regions, is generally disfavored (for the references see e.g.
\cite{DZ,L}).

With the advent of GUTs it became clear that the baryon number is not
conserved and BAU can be generated dynamically in the first instant after
big bang, while the initial baryon number of the Universe is zero or an
arbitrary non-zero value. Various mechanisms of baryogenesis were
constructed such as those involving quantum gravity, particle creation in
the gravitational field of an expanding Universe, the creation and
evaporation of black holes and monopoles (see also \cite{BCKL}), the
sufficient entropy generation in initially cold charge-asymmetric Universe,
the effects of anisotropy or inhomogeneities, the baryon number generation
in theories with integer charged quarks and so on. The reviews of these
mechanisms and references are given in \cite{DZ,L,Ba,ZN,Ko}.

The most popular type of models (which we will discuss here) was proposed
first by Sakharov \cite{S} and Kuzmin \cite{Ku}, who shoved that baryon
number generation requires three conditions:

(a). Some interactions of elementary particles must violate baryon number,
since the net baryon number of the Universe must change over time.

(b). $CP$ must be violated. This will provide an arrow for the direction of
$B$-violation and the rates of the $B$-violating processes will be not
equal: $\Gamma (i \rightarrow f) \neq \Gamma (\bar i \rightarrow \bar f)$,
where $i$ and $f$ denote the initial and final particles.

(c). A departure from thermal equilibrium must take place. Certainly, if a
symmetrical Universe is in thermal equilibrium, particle number densities
behave as $exp(-m/T)$. $CPT$ invariance guarantees that a particle and its
antiparticle have the same mass, and unitarity requires that the total
production rates of a particle and its antiparticle be equal, so their
densities remain equal during expansion and no asymmetry arises, regardless
of $B$-, $C$- and $CP$-violating interactions (see e.g. \cite{DZ,KW}).

GUTs provide (a) due to interactions of quarks and leptons with heavy bosons
(For the minimal $SU(5)$ GUT this interactions are
$X \leftrightarrow UU$,
$X \leftrightarrow \bar D\bar L$,
$Y \leftrightarrow UD$,
$Y \leftrightarrow \bar U\bar L$,
$Y \leftrightarrow \bar D\bar \nu$,
$H_t \leftrightarrow \bar U\bar D$,
$H_t \leftrightarrow UL$,
$H_t \leftrightarrow D\nu$,
where $X,Y$ are leptoquarks, $H_t$ --- Higgs color triplet, $U$ --- up-like
quarks, $D$ --- down-like quarks, $L$ --- charged leptons and we used the
notation of \cite{L}, opposite to \cite{FOT,NW}) and they can naturally
contain (b) through loop processes (see \cite{NW,P}), while (c) arises in
the early Universe, when particle reaction rates $\Gamma$ lag behind the
rate of cosmological expansion $H$.

The scenario GUT-scale creation  of BAU (see e.g.
\cite{FOT,DZ,L,KW,NW,BSW,DS,W,TTWZ}) is as follows.

Soon after big bang the baryon number violating interactions came into
equilibrium so that any initial baryon number was washed out. After that,
when the temperature fell down below the value $M_\chi$ (where $\chi \equiv
X,Y or H_t$), the decay rates of heavy bosons became less than the rate of
Universe expansion and $B$-violating reactions dropped out of equilibrium.
The subsequent decays of heavy bosons (inverse decays were blocked since
typical particles had energies $\sim T < M_\chi$) have generated the BAU.
This could have occurred despite the equal densities and the equal total
decay rates (which follow from $CPT$) of the $\chi$ and $\bar \chi$, because
if $C$ and $CP$ were violated, one could still have different partial decay
rates:
$$
\Gamma (\chi \rightarrow f) \neq \Gamma(\bar \chi \rightarrow \bar f) .
$$
After these decays baryon number was effectively conserved until today, as
rates of $B$-violating processes are very small for $T \ll M_\chi $, so this
baryon excess remained constant. In the and, as the time passed, the
charge-symmetrical part of plasma annihilated into photons, neutrinos and
anti-neutrinos, while the "surplus" particles survived and originated the
observed world.

The resulting baryon-to-photon ratio is \cite{FOT,L} :
\begin{equation}
\delta = \left( \frac{n_B}{n_\gamma }\right) _{initial}\cdot e^{-\xi A} +
\frac{b}{1+(cK)^d}\cdot \Delta B_\chi  ,
\label{4.2}
\end{equation}
where
$$
\Delta B_\chi = \sum_{f} B_f\cdot \frac{\Gamma (\chi \rightarrow f) -
\Gamma (\bar\chi \rightarrow \bar f)}{\Gamma _{total}(\chi )}
$$
is the total baryon number created through $\chi$ decays; parameter
$$
A = \left( 16\pi ^3 \frac{g_\ast }{45}\right) ^{-1/2}\cdot \frac{\alpha
M_{Pl}}{M_\chi } \sim \left( \frac{\Gamma }{H}\right)
$$
describes the value of nonequilibrium; $B_f$ is the baryon number of the
final state $f$; parameters $\xi =5.5(0.25\div 1.6)$, $b=0.03(0.01)$,
$c=16(3)$, $d=1.3(1.2)$ for $\chi \equiv XY(H_t)$; $\alpha _\chi$ is the
coupling constant;
$g_\ast = 160$ is the "effective" complete (sum over bosons and
7/8 times the sum over fermions) number of degrees of freedom in
relativistic particles.

In (\ref{4.2}) the first term describes the washout of initial asymmetry
(if such had existed), while the second term --- creation of new
asymmetry.

In the different models the $X$ and $Y$ or $H_t$ bosons were used as a
source of BAU. Some models considered washout of initial asymmetry
by $X,Y$ and consequent generation of BAU by $H_t$ bosons (see e.g.
\cite{FOT,L}).

However in the minimal $SU(5)$ model the natural choice of parameters
$M_{XY} \sim (10^{15} \div 10^{14}) GeV$, $M_{H_t} \sim 10^{13} GeV$
requires $\Delta B_{XY} \sim 10^{-5.7}$ and $\Delta B_{H_t} \sim 10^{-8.5}$,
while in this model $\Delta B_\chi < 10^{-10}$ \cite{NW,BSW}. Thus the
minimal $SU(5)$ model fails to explain the observable BAU and one needs to
consider different extended models, for example those with extra Higgses
(see also \cite{KST}).

Another difficulty of GUT-scale baryogenesis models is that in the standard
model of electroweak interactions baryon number is known theoretically to be
anomalous and not exactly conserved \cite{tH}. At low temperature this
anomalous baryon violation only proceeds via an exponentially suppressed
tunneling process, at a rate proportional to $exp(-4\pi /\alpha _w) \sim 0$,
where $\alpha_w$ is the weak coupling constant. At temperatures above $\sim
100 GeV$, however, electroweak baryon violation may proceed rapidly enough
to equilibrate to zero any baryons produced by Grand Unification \cite{M}.

In the last decade much work has been done to solve this problem.

One and the most popular field of investigations opened in 1985 by the work
of Kuzmin, Rubakov and Shaposhnikov \cite{KRS}, who showed that at the scale
of electroweak phase transitions there are all three necessary ingredients
for baryogenesis ($B$-violation via anomalous processes, $CP$-violation in
Kobayashi-Maskawa matrix, departure from thermal equilibrium if the
electroweak phase transition is of the first order) and BAU can be created
at this scale. After that many mechanisms of electroweak-scale baryogenesis
were proposed (for the review see \cite{MRTS,FS,CKN,D}).

Most of these mechanisms consider different extensions of standard model
(such as singlet Majoron model with complex phases in the neutrino mass
matrix; the nonsupersymmetric two Higgs doublet model with a $CP$-violating
relative phase between the doublets; and minimal supersymmetric standard
model with a $CP$-violating phase in the gaugino masses), since
$CP$-violation from Kobayashi-Maskawa matrix is by many orders small to give
the observed value for $\delta$ and one has to search for additional source
of $CP$-violation. Besides, there arise problems with the mass of higgs
particles (see \cite{FS,CKN}) - the schemes of baryogenesis in the
Minimal Standard Model require the Higgs mass to be less than $50 GeV$ that
is forbidden experimentally. Nevertheless, there are some attempts to
describe the BAU via electroweak-scale processes in the frames of Minimal
Standard Model \cite{FS,BG2} (some problems of the model proposed in
\cite{FS} are discussed in \cite{O}).  In \cite{BG2} a new source of
$CP$-violation comes from $P$-violation by the domain walls, appearing in
the first-order electroweak phase transitions, while the problem of higgs
mass is solved due to the mass-difference in the leptonic sector (see
\cite{DR}).

In the other set of works it was noted, that anomalous electroweak processes
conserve $B-L$, and so a net $B-L$ generated by GUT-scale baryogenesis will
not be erased by electroweak interactions. In the Minimal Standard Model if
$B-L$ is non-zero, then the equilibrium baryon number at high temperature is
$B\sim (B-L)$ \cite{D,HT}. However, in minimal GUTs $B-L = 0$, thus again
some extensions of standard models are necessary.

There are also models considering cases when the early generated BAU will
not be washed out even if $B-L = 0$. For example, in \cite{KRS-87} it was
argued, that equilibrium anomalous electroweak $B$-non-conserving
interactions do not wash all the baryon asymmetry if (1) there is no mixing
in the leptonic sector, (2) there is large flavor asymmetry in the leptonic
decays, (3) mass of the Higgs boson is larger than 56 GeV. The idea of
asymmetry in leptonic sector was used also in \cite{DR} where it was claimed
that different lepton masses provide the different leptonic chemical
potentials and a net $B$-asymmetry survives. In the work \cite{Bar} the
conclusion was made that if there is another global quantum number in the
low-energy theory that is, like $B$ and $L$, conserved up to weak anomaly,
the electroweak anomaly processes would not necessarily erase a primordial
baryon asymmetry even if $B-L=0$. There are also considerations of SUSY
effects, which can save primordial BAU \cite{IQ}.

In any case the above said shows that the question of GUT-scale baryogenesis
is still open and needs further investigation.

\section {Particle-antiparticle mass-difference and Baryogenesis in
$SU(5)$ model}

{}~~~~~~As we have discussed in previous sections, the baryon excess is usually
expected to be produced during the nonequilibrium stage of the Universe
expansion, since if $CPT$ is valid, no non-conserved quantum number (such as
baryon number) could take a nonzero value when averaged over a statistical
at thermodynamic equilibrium. Now we shall take into account that $CPT$ is
not valid due to Universe expansion and thus masses of particles and
antiparticles are slightly different. One might expect that there could
exist some excess in the number density of particles over antiparticles (or
vice-versa) in plasma even at thermal equilibrium. This could in turn result
in the non-vanishing average macroscopic value of the baryon number in the
plasma at equilibrium. We find that this is indeed the case. Let us consider
this in more detail.

Let us discuss, what does happen in cosmological plasma in thermal
equilibrium at high temperatures in the frames of some GUT theory if the
masses of leptoquarks and antileptoquarks are different. The thermal
equilibrium in an expanding Universe establishes when the particle
interaction rates are rapid compared to the expansion rate. We shall
consider the temperature interval $10^{15}GeV \sim M_{X,Y} > T \geq T_f \sim
10^{14}GeV$, where $M_{X,Y}$ is the mass of leptoquarks and $T_f \sim 0.1
M_X$ is the "freezing" temperature at which the massive leptoquarks in
plasma decouple.

As a starting point we will assume that the World is described by the
minimal SU(5) model. The particle contents of this model is as follows (see
e.g. \cite{L}):

(a). Three families of quarks and leptons. Each family (labelled by $(j)$)
contains 15 twocomponent fermions:
$$
(\nu , e^-)_{(j)L} \; : \; ({\bf 1, 2 }) ; \quad
e^+_{(j)L} \; : \; ({\bf 1, 1 }) ; \quad
(u_\alpha , d_\alpha)_{(j)L} \; : \; ({\bf 3, 2 }) ;
$$
$$
u^{c \alpha}_{(j)L} \; : \; ({\bf 3^\ast , 1 }) ; \quad
d^{c \alpha}_{(j)L} \; : \; ({\bf 3^\ast , 1 }) .
$$
Here the index $"L"$ denotes that fermions are left-handed, the index
$\alpha = 1,2,3$ corresponds to different colors of particles, the upper
index $"c"$ denotes the charge conjugation
$\left( \psi^c = C\gamma^0\psi^\ast = i\gamma^2\psi^\ast , \quad
\psi_L^c = C\bar\psi^T_R \right) $,
the entries in brackets $(n_3, n_2)$ represent the representations under the
$SU(3)^{col}$ and $SU(2)$ subgroups. These fermions can be placed in
${\bf 5}^\ast + {\bf 10}$ representation of $SU(5)$.

(b). The 24 gauge bosons: one photon
$({\bf 1}, {\bf 1})$,
8 gluons $G^\alpha_\beta$:
$({\bf 8}, {\bf 1})$,
three weak bosons $W^\pm , W^0$ :
$({\bf 1}, {\bf 3})$,
six $(X,Y)$ bosons :
$({\bf 3}, {\bf 2^\ast})$
and their six antiparticles $(\bar X, \bar Y)$ :
$({\bf 3^\ast}, {\bf 2})$ .
$X$ and $Y$ bosons have electric charge equal to $4/3$ and $1/3$
respectively.

(c). The 34 scalar bosons of the model are placed in {\bf 24} and complex
{\bf 5} representations of $SU(5)$. Out of them the physical meaning have:
the complex Higgs doublet of Weinberg-Salam model ($(H_d^+, H_d^0) \, : \,
({\bf 1}, {\bf 2})$
and $(H_d^0, H_d^-)$); the complex color triplet of Higgs fields
$H_t \, : \, ({\bf 3}, {\bf 1})$ ;
the superheavy scalar bosons placed in representation {\bf 24}:
$\Phi _\beta^\alpha \, : \,
({\bf 8}, {\bf 1})$ ;
$\Phi^r_s \, : \,
({\bf 1}, {\bf 3})$
and $\Phi \, : \,
({\bf 1}, {\bf 1})$.
The remaining twelve bosons from representation {\bf 24} are unphysical and
give masses to $X$ and $Y$ gauge bosons. Here the indices $r,s$ run through
the values $4,5$.

In $SU(5)$ there are two types of bosons that mediate $B$- and $L$-violating
interactions --- the $X$ and $Y$ gauge particles and color triplet of Higgs
scalars $H_t$. However $H$ bosons typically have a smaller effective
coupling constant
$(\alpha_H \sim 10^{-4} \div 10^{-6})$ than $X,Y$ bosons do $(\alpha \sim
10^{-2}$ (see e.g. \cite{FOT}) and here we will only consider the effect of
the $X$ and $Y$ gauge bosons on the evolution of Baryon asymmetry.

The super-heavy scalars $\Phi$ from representation {\bf 24} do not couple to
fermions and we will not consider them too. Neglecting interactions with
heavy $H_t$ and $\Phi$ scalars does not affect our calculations sufficiently.

Thus for our purposes the relevant fields are: three generations of quarks
and leptons, complex doublet of Higgs particles and gauge particles. At the
temperature interval which we are considering all particles except $X$ and
$Y$ can be treated as a massless with a very good approximation.

Since $SU(5)$ is symmetrical with respect to generations and colors, we
shall treat generation and color as additional, spinlike degeneracies
(recall that the number of spin states for the particle species $i$ are
equal to $2s_i+1$ for $m_i>0$; to $2$ for $m_i=0, s_i>0$ and to $1$ for
$m_i=0, s_i=0$, where $s_i$ denotes the spin of "i" particle and $m_i$ ---
its mass). Specifically, all left- and right-handed up-like quarks
($u_{(j)L}$ and $u_{(j)R}$)
will be referred as $u_L$ and $u_R$ with an effective degeneracy factor
$ g_{u_L} = g_{u_R} = g_u = (1$ helicity state$)\times (3$
generations$)\times (3$ colors$) = 9$;
all left- and right-handed down-like quarks
($d_{(j)L}$ and $d_{(j)R}$)
are designated $d_L$ and $d_R$ with an effective degeneracy $g_{d_L}
=g_{d_R} = g_d = 9$.
All left- and right- handed charged leptons
($e_{(j)L}$ and $e_{(j)R}$)
are labeled by $e_L$ and $e_R$ and have an effective degeneracy
$ g_{e_L} = g_{e_R} = g_e = (1$ helicity state$)\times (3$ generations$) = 3$
and finally, all neutrinos $\nu _{(j)}$ are labeled by $\nu$, where $\nu$ has
an effective degeneracy
$ g_\nu = (1$ helicity state$)\times (3$ generations$) = 3$.

The $X$ and $Y$ superheavy gauge bosons have an effective degeneracy
$ g_X = g_Y = (3$ spin states$)\times (3$ colors$) = 9$.
The $X$ and $Y$ masses are assumed to be equal and each is denoted $M_X$.
Any difference is due to $SU(2)\times U(1)$ weak symmetry breaking and is
of order 100GeV, negligible at temperatures which we are investigating.

Since we are investigating a cosmological plasma in thermal equilibrium, the
particle distributions will be given by their thermal forms. If we consider
a uniform ideal gas of particle species $"i"$ with mass $m_i$ in thermal
equilibrium at temperature $T$, then the number density of such particles in
phase space is given by (see \cite{KW})
\begin{equation}
\frac{dN_i}{d^3\vec p d^3x} \equiv f_i(p) = \frac{g_i}{(2\pi )^3}\cdot
\frac{1}{e^{(E_i - \mu_i)/T} + \theta } .
\label{4.3}
\end{equation}
Here $E_i = \sqrt{\vec p^2 + m_i^2}$, $\theta = \pm 1$ for particles obeying
Fermi-Dirac (Bose-Einstein) statistics and $\theta = 0$ in the classical
(distinguishable particles) approximation of Maxwell-Boltzmann statistics;
$g_i$ gives the effective degeneracy for particle species $"i"$ (see above).
The $\mu$ is a possible chemical potential which serves to constrain the
total number of particles.

The total particle number density may be obtained by integrating
(\ref{4.3}) over available momentum states:
\begin{flushleft}
\begin{eqnarray}
n_i(T,\frac{m_i}{T},\frac{\mu_i}{t}) = \int \limits_{0}^{\infty}
\frac{d^3\vec p}{2\pi^3}\cdot \frac{g_i}{e^{(E_i - \mu_i)/T} + \theta } =
\nonumber \\
= \frac{g_iT^3}{2\pi^3}\cdot \int \limits_{m_i/T}^{\infty}
q\sqrt{q^2 - (m_i/T)^2}\cdot \left[ e^{q-\mu_i/T} + \theta \right] ^{-1}dq .
\nonumber
\end{eqnarray}
\end{flushleft}
We shall make the simplifying approximation that all particles obey
Maxwell-Boltzmann statistics. The corrections resulting the
indistinguishibility of the particles are small (see \cite{KW}). In this
approximation the number density integral becomes:
\begin{flushleft}
\begin{equation}
n_i(T,\frac{m_i}{T},\frac{\mu_i}{t}) \simeq
g_i\cdot \left( \frac{m_iT}{2\pi}\right)^{3/2}\cdot e^{-(m_i - \mu_i)/T}
\cdot \left[ 1 + \frac{15}{8}\frac{T}{m_i} + \cdots \right] \;
(for T \ll m_i ) ;
\label{4.4}
\end{equation}
\end{flushleft}
\begin{flushleft}
\begin{equation}
n_i(T,\frac{m_i}{T},\frac{\mu_i}{t}) \simeq
g_i\cdot \left( \frac{T^3}{\pi^2}\right)\cdot e^{\mu_i/T}
\cdot \left[ 1 - \frac{1}{4}\left(\frac{m_i}{T}\right)^2 - \cdots \right] \;
(for T \gg m_i ) ;
\label{4.5}
\end{equation}
\end{flushleft}

It is easy to see from (\ref{4.4}) and (\ref{4.5}) that the excess of
particle species $"i"$ over their antiparticles can be determined via
chemical potentials. Thus now we have to discuss the relevant chemical
potentials of used model.

In thermal equilibrium any charged particle can emit or absorb an arbitrary
number of photons; thus the chemical potential of the photon vanishes at
sufficiently high temperatures and densities. An analogous conclusion holds
for the $Z^0$ boson (for $T \geq M_{Z^0}$). Therefore, as a consequence of
processes such as $\gamma\gamma \rightarrow b\bar b$, the chemical
potentials of particles and antiparticles are equal and opposite. At
temperatures above the electroweak phase transition (when $U(2)$ is an exact
symmetry) the $W^\pm$ also have a vanishing chemical potentials, what
imposes equality of the chemical potentials for fields in the same
electroweak multiplet (so that $\mu (X) = \mu (Y),\; \mu (u_L) = \mu (d_L),
\; \mu (e_L) = \mu (\nu)$). Similarly (as $SU(3)$ is an exact symmetry) the
chemical potential of gluon vanishes and thus the different colored quarks
have equal chemical potentials. Moreover, Cabbibo mixing will guarantee that
chemical potentials of all up-like quarks and all down-like quarks are
respectively equal ($\mu (u_{(j)}) = \mu (u); \; \mu (d_{(j)}) = \mu (d)$).
In the absence of flavor mixing neutrino processes (e.g. due to neutrino
  masses), the lepton generations will not in general have an equal
  chemical potentials, but as we shall see soon, interactions with
  leptoquarks and quarks ensure that leptons of all generations have equal
  chemical potentials. However, the chemical potentials of left particles,
  in general, do not equal to those of right particles.  In all, there are
  7 different chemical potentials which we have to calculate.
  Fortunately thermodynamical equilibrium imposes a number of relations
  between them.

Recall, that whenever interactions are in thermal equilibrium, the sum of
the chemical potentials of the incoming particles is equal to that of
outgoing particles (see \cite{LL}). Rapid electroweak interactions in the
early Universe enforse the following equilibrium relations among the
chemical potentials (see \cite{HT,DR} and \cite{FOT,L,NW}):
\begin{flushleft}
\begin{eqnarray}
(a).\, \mu (u_R) = -\mu (u^c_L) = -\mu (u_L) + \mu (X)
\quad  ( u_L \leftrightarrow u^c_L + X ) ; \nonumber \\
(b).\, \mu (H_d^0) = -\mu (u_L) + \mu (u_R) = -2\mu (u_L) + \mu (X)
\quad  ( H_d^0 \leftrightarrow \bar u_L + u_R) ; \nonumber \\
(c).\, \mu (d_R) = -\mu (H_d^0) + \mu (u_L) = 3\mu (u_L) - \mu (X)
\quad  (H_d^0 \leftrightarrow d_L + \bar d_R) ; \nonumber \\
(d).\, \mu (e_{(j)R}) = -\mu (e_{(j)L}^+)  = - \mu (u_L) - \mu (X)
\quad  (X + d_L \leftrightarrow e_{(j)L}^+) ; \nonumber \\
(e).\, \mu (e_{(j)L}) = -\mu (H_d^0) + \mu (e_{(j)R})  = - 3\mu (u_L)
\quad  (H_d^0 \leftrightarrow e_{(j)L}^ + \bar e_{(j)R}) .
\label{4.6}
\end{eqnarray}
\end{flushleft}

The other reactions are correlated with these by $SU(2)$ symmetry.

We see, that by the use of these relations all of the chemical potentials
  can be expressed by means of $\mu (u_L)$ and $\mu (X)$.

We have also one more unknown variable $\Delta m/m$.

  To evaluate this remaining quantities one can use the restrictions coming
  from conservation of electric charge, color, weak isospin and $B-L$. I.e.
  the cosmological plasma must be neutral with respect to charge, color,
  weak isospin and $B-L$. The neutrallness with respect to color and weak
  isospin is satisfied automatically, since at the temperature interval,
  which we are considering, $SU(3)^{col}$ and $SU(2)$ are exact symmetries.
  The conditions of charge and $B-L$ neutrallness can be expressed in terms
  of particle densities in plasma and have the form:
  \begin{eqnarray}
  Q =
  \frac{4}{3}\left[ n(X) - n(\bar X) \right] -
  \frac{1}{3}\left[ n(Y) - n(\bar Y) \right] +
  \frac{2}{3}\left[ n(u) - n(\bar u) \right] - \nonumber \\
  \frac{1}{3}\left[ n(d) - n(\bar d) \right] -
  \left[ n(e) - n(e^+) \right] -
  \left[ n(W^-) - n(W^+) \right] +
  \left[ n(H_d^+) - n(H_d^-) \right] = 0
  \label{4.7}
  \end{eqnarray}
  \begin{eqnarray}
  B-L =
  \frac{2}{3}\left[ n(X) - n(\bar X) \right] +
  \frac{2}{3}\left[ n(Y) - n(\bar Y) \right] +
  \frac{1}{3}\left[ n(u) - n(\bar u) \right] + \nonumber \\
  \frac{1}{3}\left[ n(d) - n(\bar d) \right] -
 - \left[ n(e) - n(e^+) \right] -
  \left[ n(\nu ) - n(\bar \nu) \right] = 0
  \label{4.8}
  \end{eqnarray}

  From (\ref{4.4}) and (\ref{4.5}) it is easy to see, that the difference
  between the densities of given particle species and the densities of their
  antiparticles can be expressed via their chemical potentials.

  We shall assume that $X(Y)$ and $\bar X(\bar Y)$ have the different masses
\begin{eqnarray}
  M_{X(Y)} = m + \Delta m ; \nonumber \\
  M_{\bar X(\bar Y)} = m - \Delta m .\nonumber
  \end{eqnarray}
  and use for their densities formula (\ref{4.4}). Then, assuming that
  $\delta m/m \ll 1, \; \mu /T \ll 1, \; \Delta m/T \sim 0.1 \ll 1$,
  we yield (hereafter, if not necessary, we shall write simply $X$ instead
  of $X,Y$):
  \begin{eqnarray}
  n(X) - n(\bar X) \simeq
g_X\cdot \left( 2\frac{T^3}{\pi^2}\right) \cdot \left( \frac{\sqrt{\pi}}{2
\sqrt{2}}\right) \cdot
\left[ \left( \frac{m}{T} \right) ^{3/2} \cdot e^{-m/T}\right] \cdot
\left( \frac{\mu (X)}{T} - \frac{\Delta m}{T} - \frac{3}{8}\frac{\Delta m}{m}
\right)
\approx \nonumber \\ \approx
g_X\cdot \left( 2\frac{T^3}{\pi^2}\right) \cdot 8.9\cdot 10^{-4} \cdot
\left( \frac{\mu (X)}{T} - \frac{\Delta m}{T}  \right)
   \label{4.9}
  \end{eqnarray}
  For all the other particles we must use formula (\ref{4.5}) since their
  masses are zero or much less than $T$. Then we yield:
  \begin{equation}
  n(i) - n(\bar i) = g_i\cdot\left( 2\frac{T^3}{\pi ^2}\right) \cdot
  \frac{\mu (i)}{T}
  \label{4.10}
  \end{equation}

  Substituting this formulae into plasma neutrallness conditions
  (\ref{4.7}) and (\ref{4.8}) we yield two additional equations
  establishing correlations between the chemical potentials and $\Delta
  m/m$, which after using the relations (\ref{4.6}) take the form (we
  have divided them by the factor $2T^3/\pi ^2$):
  \begin{eqnarray}
  Q = -5\cdot (2.67\cdot 10^{-3})\cdot \frac{\Delta m}{T} - 2\cdot
\frac{\mu (u_L)}{T} + 13\cdot \frac{\mu (X)}{T} = 0  \nonumber \\
  B-L = -4\cdot (2.67\cdot 10^{-3})\cdot \frac{\Delta m}{T} + 33\cdot
\frac{\mu (u_L)}{T} + 3\cdot \frac{\mu (X)}{T} = 0  \nonumber
  \end{eqnarray}

  Solving this system of equations gives
  \begin{eqnarray}
  \mu (u_L) \simeq 0.21 \mu (X) \nonumber \\
  \mu (X) \simeq 1.07\cdot 10^{-3}\Delta m.
  \label{4.11}
  \end{eqnarray}

  Now we can express the particle-over-antiparticle excess for leptoquarks
and quarks via the mass difference $\Delta m$. This is necessary to find the
net baryon number carried by leptoquarks and quarks. Using (\ref{4.9}),
(\ref{4.10}) and (\ref{4.11}) we yield:
\begin{equation}
\frac{n(X) + n(Y) - n(\bar X) - n(\bar Y)}{n_\gamma} \simeq (g_X + g_Y)
\cdot 8.9 \cdot 10^{-4}\cdot \left( \frac{\mu (x)}{T} - \frac{\Delta m}{T}
\right) \simeq -1.6\cdot 10^{-1} \cdot \frac{\Delta m}{m}
\label{4.12}
\end{equation}
\begin{equation}
\frac{n(q) - n(\bar q)}{n_\gamma} \simeq
g_{u_L}\cdot \frac{\mu (u_L)}{T} +
g_{u_R}\cdot \frac{\mu (u_R)}{T} +
g_{d_L}\cdot \frac{\mu (d_L)}{T} +
g_{d_R}\cdot \frac{\mu (d_R)}{T}
\simeq 0.8\cdot 10^{-1} \cdot \frac{\Delta m}{m}
\label{4.13}
\end{equation}
Thus for the baryon number excess we have
  $$
\frac{\Delta B_{X,Y}}{n(\gamma )} = \frac{1}{6}\frac{\Delta n(X)}{n(\gamma)}
\simeq -0.25 \cdot 10^{-1}\cdot \frac{\Delta m}{m}
  $$
  $$
\frac{\Delta B_{q}}{n(\gamma )} = \frac{1}{n(\gamma)}(\Delta n(u) + \Delta
n(d)) \simeq 0.8\cdot 10^{-1}\cdot \frac{\Delta m}{m}
  $$
Here $\Delta B_{X,Y}$ and $\Delta B_{q}$ is the net baryon number carried by
leptoquarks and quarks respectively; the factor 1/6 in (\ref{4.12}) is the
mean excess of baryon number in decays of single leptoquark in vacuum (see
\cite{FOT}); for $n(\gamma )$ we have used formula (\ref{4.5}) with
$m_{\gamma} = 0, \mu (\gamma ) = 0$.

Using (\ref{4.12}) and (\ref{4.13}) one can see that the complete BAU is
  \begin{equation}
\delta \simeq \frac{1}{g_\ast} \left( \frac{\Delta B_{X,Y}}{n_{\gamma}} +
\frac{\Delta B_q}{n_{\gamma}} \right) \simeq 0.55\cdot 10^{-3} \frac{\Delta
  m}{m} .
\label{4.14}
  \end{equation}

Thus we have found that if there exists the mass difference between
leptoquarks and antileptoquarks then in primordial plasma there will be the
excess of baryons over antibaryons even in thermal equilibrium. Comparing
this result with observed data one can conclude that under the assumption
that BAU is completely caused by $CPT$-violation this mass difference must
be of the order of
  $$
\frac{\Delta m}{m} \sim 10^{-4} \div 10^{-6} .
  $$

\section{Calculation of BAU caused by the Universe expansion}

{}~~~~~As we  have already discussed, the standard $SU(5)$ GUT fails to
explain the observable BAU since the lack of sufficient nonequilibrium in
this model. In previous section it was shown that if one takes into account
that $CPT$ theorem is violated on GUT scale and causes the mass difference
between the leptoquarks and antileptoquarks
\begin{equation}
\Delta m \sim (10^{-4} \div 10^{-6})m ,
\label{4.15}
\end{equation}
one can obtain the observable value of BAU without the requirement of
nonequilibrium.

Now let us show that $CPT$-invariance in FRW space is really violated and on
GUT scale it causes the mass difference (\ref{4.15}).

  Our task is to find the quantity $\Delta m_X/m_X$. We have already
 obtained that in RW universe  the mass difference obtained between
particles and their antiparticles via Higgs mechanism is
\begin{equation}
\frac{\Delta m_X}{m_X} = (\dot a|_{\tau = 0})\tau /a_0 .
\label{4.16}
\end{equation}

Let us consider the case of leptoquarks. As the Universe cools down the
first phase transition takes place at the GUT scale at the time moment
  $$
t \sim \frac{1}{M_X} .
  $$
At this moment the group of GUT, for example $SU(5)$, breaks down and the
leptoquarks acquire their masses. All the other particles of the model
are massless at this scale. Let us assume that the Leptoquarks acquire
their masses during a sufficiently short period
  $$
 t \sim a_0\tau \sim (1 \div 10^{-1}) / m_X , \quad \tau \ll 1 .
  $$
Then we can use the short-time-interval approximation and so the
expression (\ref{4.16}) is valid.

Let us evaluate this value for the different expansion regimes \cite{BG1} .

  As we have seen, during the period when $\rho = const$ the evolution of
the Universe is described (according to (\ref{2.8})) by the de  Sitter
expansion law
\begin{equation}
a = a_0\cdot exp(H t) = a_0 /(1 - a_0H\tau  )
\label{4.17}
\end{equation}
with the Hubble constant
$$H =  K\sqrt{\rho},$$
where the constant $K$ is expressed by (\ref{2.7})

  The energy density $\rho$ of a nonlinear scalar field leading to
inflation in some models is  of the order of
$$\rho \sim (10^{-2} \div 1) m_X^4  .$$
{}From (\ref{4.17}) it follows that
$$\dot a |_{\tau =0} = a_0^2H $$
and according to (\ref{4.16}) we obtain:
\begin{equation}
\frac{\Delta m_X}{m_X} \sim H a_0\tau \sim Ht
\sim (10^{-2} \div 1)\cdot\frac{m_X}{M_{pl}} .
\label{4.18}
\end{equation}
If the mass of the leptoquarks is generated at the very end of the
inflationary period, then, taking into account that
$m_X / M_{pl} \sim 10^{-4}$,
eq. (\ref{4.18}) gives the value (\ref{4.15}), required to explain the
BAU.

    When the ultrarelativistic  gas  of  noninteracting  particles
dominates in the Universe $(\rho  = a^{-4})$,  then  according  to
(\ref{2.9}) the scale factor depends on time through the power-function
law:
$$
 a = (a_0^2 + 2Kt)^{1/2} =  a_0 + K\tau .
$$
 Then
$$a|_{\tau = 0} = K $$
and
\begin{equation}
\frac{\Delta m_X}{m_X} \sim \frac{K}{a_0}\tau \sim
\frac{K}{a_0^2}t.
\label{4.19}
\end{equation}
If leptoquarks acquire their mass during this period and the  value
of scale factor at the initial moment of phase transition is of the
order of
$$a_0 \sim (1 \div 10)/m_X, $$
then we obtain (\ref{4.15}) again.  Note that at the
power-function-like expansion of the Universe the mass difference between
particles and antiparticles is proportional to $a_0^{-2}$ (see
(\ref{4.19})) and for the particles acquiring their mass  later (at large
$a_0$) it tends to zero.  This result is valid also for the expansion law
$a\sim t^{2/3}$.  Thus the $CPT$-violation causing the mass difference
between the particles and antiparticles is essential only during the
early stages of the universe expansion.

    We have shown that for a wide spectrum of models the  observed
BAU can be explained by the violation of the $CPT$ -theorem in the
strong gravitational fields at the early stages of the Universe expansion
without requirement of nonequilibrium and  additional  complication  of
model. Besides, as we have seen in subsection 3.3, in the expanding universe
$\Delta m_X$ is positive and according to (\ref{4.14}) we have the
predominance of matter over antimatter. In the collapsing universe $\Delta
m_X$ is negative and we should observe the opposite picture. Thus the
predominance of matter over antimatter and the expansion of the Universe
seem to be connected facts.  This conclusion is valid also with respect to
the other models of GUT (besides of $SU(5)$) though the numerical
evaluations can be different.

        At the end we would like to note that to discern the properties of
particles and antiparticles in Minkowski space it is necessary to have the
$C$-noninvariant interaction. However in the curved space it is not so if
$CPT$ is violated. As we have sown, the different behavior of particles and
antiparticles can be caused by a cosmological $T$-asymmetry and thus BAU can
be generated without $C$-violation.


\end{document}